\documentclass[12pt,preprint]{aastex}

\newcommand{\sn}{SN~1987A}
\newcommand{\hst}{{\it HST\/}}
\newcommand{\iue}{{\it IUE\/}}
\newcommand{\Lya}{${\rm Ly}\alpha$}
\newcommand{\Ha}{${\rm H}\alpha$}
\newcommand{\Hb}{${\rm H}\beta$}
\newcommand{\Hg}{${\rm H}\gamma$}
\newcommand{\kms}{\rm ~km~s^{-1}}
\newcommand{\ergs}{\rm\,ergs~s^{-1}}
\newcommand{\ergcms}{\rm\,ergs~cm^{\small -2}~s^{\small -1}}

\newcommand{\wl}{$\lambda$}
\newcommand{\wll}{$\lambda\lambda$}
\newcommand{\pcc}{\mbox{ cm}^{-3}}
\newcommand{\col}{\mbox{ cm}^{-2}}
\newcommand{\K}{\mbox{ K}}

\newcommand{\hii}{{\ion{H}{2}}}
\newcommand{\hei}{{\ion{He}{1}}}
\newcommand{\heii}{{\ion{He}{2}}}
\newcommand{\cii}{{\ion{C}{2}}}

\newcommand{\civ}{{\ion{C}{4}}}
\newcommand{\nI}{{\ion{N}{1}}}
\newcommand{\nii}{{\ion{N}{2}}}

\newcommand{\niv}{{\ion{N}{4}}}
\newcommand{\nv}{{\ion{N}{5}}}

\newcommand{\oi}{{\ion{O}{1}}}
\newcommand{\oii}{{\ion{O}{2}}}
\newcommand{\oiii}{{\ion{O}{3}}}
\newcommand{\oiv}{{\ion{O}{4}}}

\newcommand{\sii}{{\ion{S}{2}}}
\newcommand{\siii}{{\ion{S}{3}}}
\newcommand{\SiII}{{\ion{Si}{2}}}

\newcommand{\SiIV}{{\ion{Si}{4}}}
\newcommand{\mgii}{{\ion{Mg}{2}}}

\newcommand{\feii}{{\ion{Fe}{2}}}

\newcommand{\amucc}{\mbox{ amu cm}^{-3}}
\newcommand{\yrs}{\mbox{ yrs}}
\newcommand{\cm}{\mbox{ cm}}
\newcommand{\lae}{\mathrel{<\kern-1.0em\lower0.9ex\hbox{$\sim$}}}
\newcommand{\pres}{\mbox{ dyne cm}^{-2}}

\newcommand{\gae}{\mathrel{>\kern-1.0em\lower0.9ex\hbox{$\sim$}}}

\begin{document}

\title{
Modelling the {\bf \it Hubble Space Telescope\/} Ultraviolet and
Optical Spectrum of Spot 1 on the Circumstellar Ring of SN~1987A
\altaffilmark{1}
}

\author{Chun S. J. Pun\altaffilmark{2,3,4},
Eli Michael\altaffilmark{5},
Svetozar A. Zhekov\altaffilmark{5,6},
Richard McCray\altaffilmark{5},
Peter M. Garnavich\altaffilmark{7},
Peter M. Challis\altaffilmark{8},
Robert P. Kirshner\altaffilmark{8},
E. Baron\altaffilmark{9},
David Branch\altaffilmark{9},
Roger A. Chevalier\altaffilmark{10},
Alexei V. Filippenko\altaffilmark{11},
Claes Fransson\altaffilmark{12},
Bruno Leibundgut\altaffilmark{13},
Peter Lundqvist\altaffilmark{12},
Nino Panagia\altaffilmark{14},
M. M. Phillips\altaffilmark{15},
Brian Schmidt\altaffilmark{16},
George Sonneborn\altaffilmark{3},
Nicholas B. Suntzeff\altaffilmark{17},
Lifan Wang\altaffilmark{18},
and J. Craig Wheeler\altaffilmark{19}}

\altaffiltext{1}{Based on observations with the NASA/ESA {\it Hubble
Space Telescope}, obtained at the Space Telescope Science Institute,
which is operated by the Association of Universities for Research in
Astronomy Inc., under NASA Contract NAS5-26555.}
\altaffiltext{2}{Laboratory for Astronomy and Space Physics, Code 681,
NASA-GSFC, Greenbelt, MD 20771.}
\altaffiltext{3}{National Optical Astronomical Observatories,
P.O. Box 26732, Tucson, AZ 85726.}
\altaffiltext{4}{Present address: Dept. of Physics, University
of Hong Kong, Pokfulam Road, Hong Kong; jcspun@hkucc.hku.hk.}
\altaffiltext{5}{JILA, University of Colorado, Boulder, CO 80309-0440.}
\altaffiltext{6}{On leave from Space Research Institute,
Sofia, Bulgaria.}
\altaffiltext{7}{Dept.\ of Physics,
University of Notre Dame, 225 Nieuwland Science Hall, Notre Dame, IN 46556.}
\altaffiltext{8}{Harvard-Smithsonian Center for Astrophysics, 60
Garden St, Cambridge, MA 02138.}
\altaffiltext{9}{Dept.\ of Physics and Astronomy, University of Oklahoma,
440 W. Brooks St., Norman, OK 73019.}
\altaffiltext{10}{Dept.\ of Astronomy, University of Virginia, P.O. Box
3818, Charlottesville, VA 22903.}
\altaffiltext{11}{Dept.\ of Astronomy, University of California,
Berkeley, CA 94720-3411.}
\altaffiltext{12}{SCFAB, Stockholm Observatory, Dept. of Astronomy,
SE-10691 Stockholm, Sweden.}
\altaffiltext{13}{European Southern Observatory, Karl-Schwarzschild-Strasse 2,
D-85784 Garching, Germany.}
\altaffiltext{14}{STScI, 3700 San Martin Drive, Baltimore, MD 21218; on
assignment from the Space Science Department of ESA.}
\altaffiltext{15}{Carnegie Institution of Washington, Las Campanas Obs,
Casilla 601, Chile.}
\altaffiltext{16}{Mount Stromlo and Siding Spring Observatories, Private Bag,
Weston Creek P. O., ACT 2611, Australia.}
\altaffiltext{17}{Cerro Tololo Inter-American Observatory,
Casilla 603, La Serena, Chile.}
\altaffiltext{18}{Institute for Nuclear and Particle
Astrophysics, E. O. Lawrence Berkeley National Lab, Berkeley, CA 94720.}
\altaffiltext{19}{Dept.\ of Astronomy, University of Texas, Austin, TX 78712.}

\begin{abstract}

We report and interpret \hst/Space Telescope Imaging Spectrograph
(STIS) long-slit observations of the optical and ultraviolet (UV)
($1150 - 10270$~\AA) emission-line spectra of the rapidly brightening
Spot~1 on the equatorial ring of SN 1987A between 1997 September and
1999 October (days~3869~-- 4606 after outburst).  The emission is
caused by radiative shocks created where the supernova blast wave
strikes dense gas protruding inward from the equatorial ring.  We
measure and tabulate line identifications, fluxes and, in some cases,
line widths and shifts.  We compute flux correction factors to account
for substantial interstellar line absorption of several emission
lines.

Nebular analysis shows that optical emission lines come from a region
of cool ($T_e \approx 10^4 \K$) and dense ($n_e \approx 10^6 \pcc$)
gas in the compressed photoionized layer behind the radiative shock.
The observed line widths indicate that only shocks with shock
velocities $V_s < 250 \kms$ have become radiative, while line ratios
indicate that much of the emission must have come from yet slower
($V_s \lae 135 \kms$) shocks.  Such slow shocks can be present only if
the protrusion has atomic density $n \gae 3 \times 10^4 \pcc$,
somewhat higher than that of the circumstellar ring.  We are able to
fit the UV fluxes with an idealized radiative shock model consisting
of two shocks ($V_s = 135$ and $250 \kms$).  The observed UV flux
increase with time can be explained by the increase in shock surface
areas as the blast wave overtakes more of the protrusion.  The
observed flux ratios of optical to highly-ionized UV lines are greater
by a factor of $\sim 2 - 3$ than predictions from the radiative shock
models and we discuss the possible causes.  We also present models for
the observed \Ha\ line widths and profiles, which suggests that a
chaotic flow exists in the photoionized regions of these shocks. We
discuss what can be learned with future observations of all the spots
present on the equatorial ring.

\end{abstract}


\section{Introduction}

Supernova 1987A (\sn) in the Large Magellanic Cloud provides an
unprecedented opportunity to observe the birth and development of a
supernova remnant.
{\it International Ultraviolet Explorer\/} (\iue) observations
\citep{fran89,sonn97} found narrow line emission about 80 days after the
explosion\footnote{The time of core collapse of \sn, 1987 February 23.316
(UT) (JD = 2,446,849.816), was determined by the IMB and Kamiokande II
neutrino detectors \citep{bion87,hira87}.},
demonstrating the presence of circumstellar gas around \sn.
Images taken with the {\it Hubble Space Telescope\/} (\hst) showed
that this gas consists of an
equatorial inner ring (radius $\approx 0.7$~light year, $n_e \sim 3\times
10^3 - 3\times10^4 \pcc$) and two outer rings ($\sim 3$ times the size
of the inner ring, $n_e \lae 2000\pcc$) tilted towards the observer
at $\sim 45\arcdeg$ \citep{jako91,wang91,plai95,burr95,lund96,mara00,lund01}.
The circumstellar ring system was excited by the ionizing radiation
from the supernova during the shock breakout \citep{lund91,ensm92,blin00}.
In the interacting winds model, the \sn\ ring system is part of a
bipolar shell around the supernova \citep{blon93, mart95, link01}.

The first signal of ongoing interaction between the \sn\ debris and
the circumstellar gas was the rebirth of the supernova in X-ray
(Beuermann, Brandt, \&~Pietsch 1994; Gorenstein, Hughes, \&~Tucker 1994;
Hasinger, Aschenbach, \&~Tr\"{u}mper 1996) and radio 
\citep{stav92,stav93} wavelengths around day~1000.  
The size of the radio-emitting region indicated that the
supernova debris expanded unimpeded at velocity $\gae 35,000\kms$ up to
$\sim$~day 1000 before slowing down to $\approx 3500 \kms$ by the
interaction \citep{gaen97,manc01}.  The X-ray and radio observations
have been explained as the interaction of the supernova ejecta with
a rather dense ($n_H \approx 100\pcc$) \hii\ region that separates the
shocked stellar wind of the supernova progenitor from the denser gas
of the inner ring \citep{chev95,bork97a}.

The ``main event'' of the birth of the Supernova Remnant~1987A
(SNR~1987A)~--- the interaction
between the supernova debris and the circumstellar rings~--- has been
anticipated since the discovery of the circumstellar gas. Predictions
of the time of the first contact ranged from 2003 \citep{luo91} to
$1999 \pm 3$ (Luo, McCray, \&~Slavin 1994) 
to $2005 \pm 3$ \citep{chev95}.  There have
been previous model predictions of radiation from this impact in X-rays
\citep{suzu93,masa94,bork97b} and in the UV/optical \citep{luo94}.  
The first definitive sign of impact between the supernova blast
wave and the inner ring was detected in the 1997 April \hst/Space Telescope 
Imaging Spectrograph (STIS) spectral images as a blueshifted 
($\sim -250\kms$) \Ha\ feature at position angle (PA) $\approx 30\arcdeg$ 
of the ring \citep{sonn98}.  Subsequent analysis of the \hst/Wide Field
and Planetary Camera 2 (WFPC2) images taken in 1997 July showed that 
a point emission, located in projection at $88\%$ of the distance 
to the ring, was increasing in brightness over a wide range of wavelengths
\citep{garn97,garn01} and could be traced back to as early as
1995 March (day~2932) \citep{lawr00a}.
The position of the brightening spot, located just inside the edge of the
inner ring, suggested that this is the result of an interaction
of the supernova blast wave with an inward protrusion of the ring.  
This brightening spot, named Spot~1, has increased in flux by a
factor of $\sim 10$ between 1996 and 1999 \citep{garn01}.
With a number of new spots ($\sim10$ in 2000 November) since then
detected all around the inner circumstellar ring
\citep{garn00,lawr00a}, we are now witnessing the full birth
of SNR~1987A.

We present here \hst/STIS UV and optical spectroscopy of Spot~1
taken by the Supernova INtensive Study (SINS) collaboration
up to day~4606 (1999 October 7).
Results from an earlier (day~3869, 1997 September 27)
STIS spectrum of Spot~1 have been presented in \citet{mich00}.
We describe the new observations and data reduction in
\S\ref{sec-obs} and report the results in \S\ref{sec-data}.
In \S\ref{uv} we describe a detailed
method to determine the intrinsic fluxes and widths of a few UV lines,
including the \SiIV\ \wll1394, 1403 and \civ\ \wll1548, 1551 doublets, 
which are strongly affected by interstellar line absorption along 
the line of sight to the supernova.

Our working model (Figure~\ref{fig:introhydro}) for Spot~1,
as well as the other spots, is that it is caused by the impact
of the supernova blast wave with a dense inward protrusion
of the ring \citep{mich00}.
When the blast wave overtakes such an obstacle, slower shocks are
transmitted into it.
Since a range of densities is present in the ring, we expect
(\S\ref{sec-globalhydro}) that the transmitted shocks will have a
range of velocities ($V_s \sim 100$ -- $1000 \kms$).  
Not all of these shocks are responsible for the observed UV/optical
emission from Spot~1 though.  
While some UV and optical line emission is produced right at 
the shock front, a
much larger amount is produced if the shocked gas undergoes thermal
collapse, i.e. the shock becomes radiative (\S\ref{sec-shockstruc}).
The time it takes for a shock to make this transition increases with
its speed and decreases with its pre-shock density.
Once the post-shock gas collapses, a shock becomes an extremely
efficient radiator of UV and optical emission lines
(\S\ref{sec-emissionlines}).
Therefore, while the range of possible velocities present in
the ring is large, we are only observing the shocks which are
slow and/or dense enough to have become radiative.
\citet{mich00} confirmed this general picture and
found that the observed line widths and line intensity ratios
indicated the emission was formed by radiative shocks 
in the velocity range $\sim 100 - 250\kms$ entering into a
dense ($n_0 \gae 10^4 \pcc$) gas.

Nebular analysis of the observed emission lines of Spot~1
(\S\ref{sec-nebanal}) suggests that the emission comes from a
region of high density ($n_e > 10^6 \pcc$),
This result confirms our picture that this emission
comes from radiative shocks, 
which can compress the pre-shock gas by a factor $\gae 100$.
In \S\ref{sec-shockmodels}, we model the UV line fluxes from
Spot~1 between day~3869 and day~4587 with the one-dimensional
steady state shock code of \citet{cox85}.
The observed UV fluxes are fitted well with models
containing emission from two radiative shocks with 
$V_s = 135$ and $250 \kms$.
We propose two scenarios to interpret the observed increase of
UV flux with time of Spot~1.
In the first scenario the density of the obstacle is low enough
so that the cooling times of shocks entering the obstacle are
comparable to the age of the spot.
The increase in UV line fluxes are then due primarily to the 
aging of the shocks as they develop thermally collapsed layers.
In the second scenario, the obstacle is dense enough so that
all the shocks cool almost immediately.
The increased observed fluxes are instead attributed 
to an increasing surface area of shock interaction.
While the second scenario fits the observed fluxes better,
we suspect that the actual light curves probably manifest a
combination of both effects.
In \S\ref{sec-lineprof} we discuss the observed line profiles and
present simulated \Ha\ line profiles based on simple geometric
models for the shock interaction.
In \S\ref{sec-disc} we discuss what we have learned by comparing
results of simplified shock models with the spectral observations,
and describe how future observations may elucidate some
unsolved problems.  
We summarize the main results in \S\ref{sec-conclusion}.

\section{Observations and Data Reduction} \label{sec-obs}

The STIS observations of Spot~1 in both optical and UV wavelengths
obtained by the SINS team are listed in Table~\ref{tb-obs}.  \sn\ is
located in a densely populated region of the LMC and appears to belong
to a loose, young cluster region \citep{pana00}.  Target acquisition
was complicated by the stars present near the supernova, especially Star~3,
a Be star of $V \sim$ 16, at 1\farcs63 away and PA = 118\arcdeg, and
Star 2, a B2 III star of $V$ = 15.0, at 2\farcs91 away and PA =
318\arcdeg \ from the supernova \citep{walb93,scud96}.  We decided to peak-up
on the nearby star S2 \citep{walk90} and offset the telescope to center
the aperture on Spot~1.  We measured the required offset from the
WFPC2 images \citep{garn01}.  Due to the uncertainties in this
measurement, Spot~1 was located at 0\farcs08 off the center of the
slit for all observations taken before 1999 August.  We reduced these
data using the standard calibration files which assumed that the
object was located at the center of the slit.  We estimate that the
offset from the center of the slit will cause us to underestimate the
measured flux by $\lae5\%$ for the 0\farcs2 data, and by $\lae0.1\%$
for the 0\farcs5 data.

With each grating setting, we took multiple ($3-5$) observations
centered at dithered positions 0\farcs5 apart along the slit.  Cosmic
rays (in the case of optical data) and hot pixels (in optical and UV
data) were removed simultaneously when the dithered raw images were
combined with the CALSTIS software developed by the STIS Investigation
Definition Team at the Goddard Space Flight Center\footnote{CALSTIS
Reference Guide,
http://hires.gsfc.nasa.gov/stis/software/doc\_manuals.html}.
Previous narrow-slit STIS spectra processed by the SINS team
with this software showed that the flux calibration of far-UV (G140L)
and near-UV (G230L) data agree to $\lae 2\%$ for the overlapping region,
while the agreement between near-UV (G230L) and optical (G430L) data
is good to $\lae 5\%$ \citep{baro00,lent01}.

\begin{figure}
\plotone{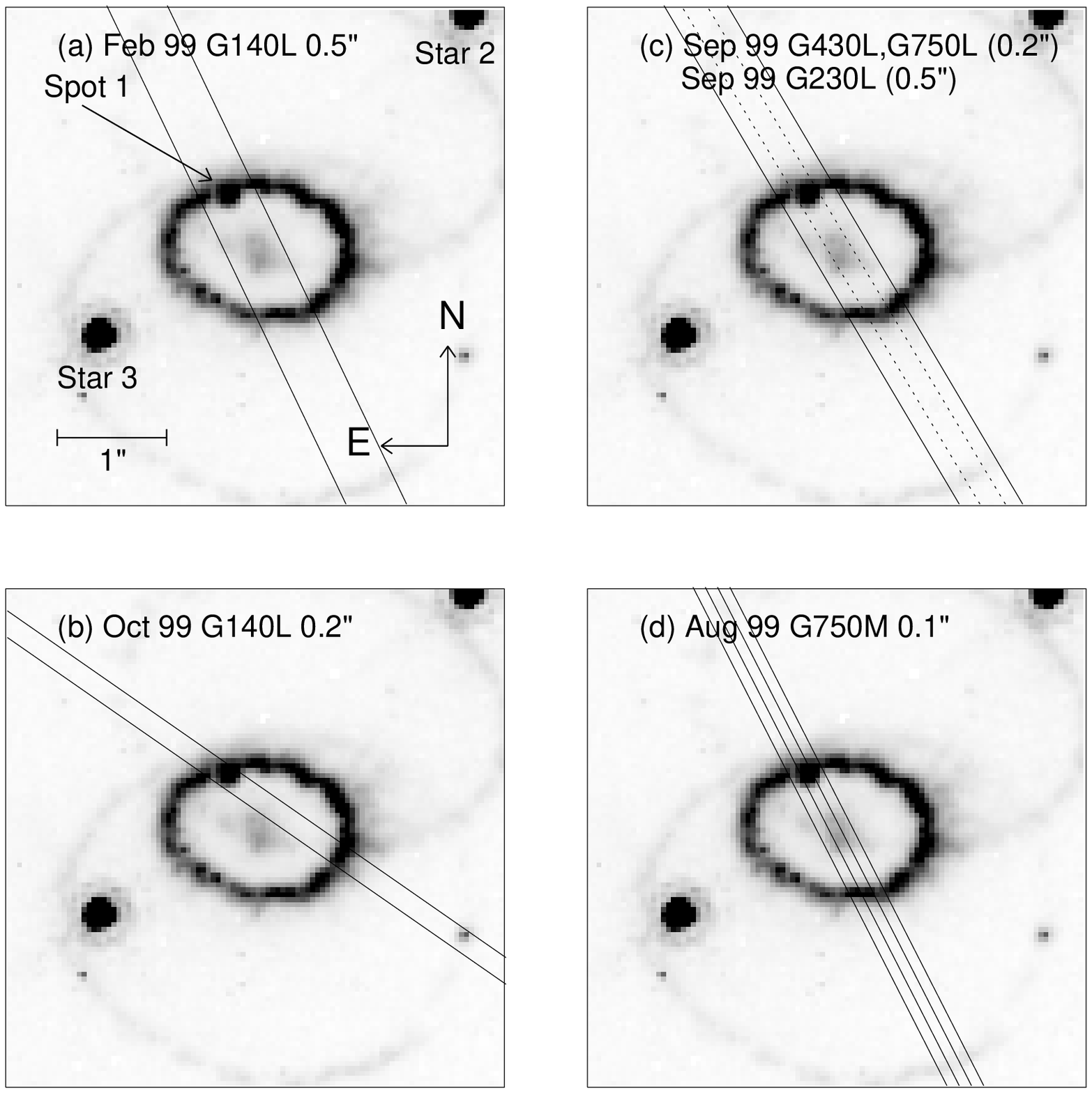}
\caption{
(a) Slit position of 1999 February 27 G140L 0\farcs5 observation
plotted on the 1999 April 21 WFPC2 \Ha\ image of \sn.
The image displayed is a $4\farcs56 \times 4\farcs56$ square.
(b) Same as (a), for the 1999 October 7 G140L 0\farcs2
observation.
(c) Same as (a), for the 1999 September 17 G230L observation
({\it solid}) and for the 1999 September 3 G430L and 1999 September 18 G750L
0\farcs2 observations ({\it dotted}).
(d) Same as (a), for the 1999 August 30 G750M 0\farcs1
observations. All three 0\farcs1 slit positions are plotted.
}
\label{fg-slitpos}
\end{figure}

The location and orientation of the aperture positions are shown
in Figure~\ref{fg-slitpos}. For all
but one of the observations, the slit was oriented (within $\pm
5\arcdeg$) along the axis connecting the center of the \sn\ debris and
Spot~1, which is located at a PA$= 29\arcdeg$ on the
inner ring \citep{garn01}.  The only exception was the
1999 October G140L observation (Figure~\ref{fg-slitpos}b),
where the ($52\arcsec \times 0\farcs2$) slit had a PA 
of 55$\arcdeg$ and did not pass through the center of the \sn\ debris.
In all observations, the Spot~1 spectrum overlapped with that 
from the adjacent segment of the inner circumstellar ring that 
was included within the STIS aperture.  With an 
expansion velocity of $10.5 \pm 0.3 \kms$ \citep{cl97,crot00}, the ring
was not resolved spectrally in any of the STIS observations reported
here.  In the optical data, the emission from the ring is the main
source of background to the Spot~1 spectrum and will be discussed in
\S\ref{opt-low} and \S\ref{opt-med}.

\citet{garn01} measured the width of Spot~1 in WFPC2 images up to 1999
April 21 (day~4440) and showed that Spot~1 was unresolved in the data
and was consistent with a point source at optical wavelengths.  We
compared the Full Width at Half-Maximum (FWHM) of Spot~1 emissions 
in our last STIS observations at 1999 September with those of point 
sources that were recorded in the data.  
We found that the FWHM of Spot~1 was $8\pm6\%$ and $25\pm9\%$
larger than a point source in the far-UV and optical wavelengths, respectively.  
The latter result is consistent with measurements by
\citet{lawr00b} with the 2000 May 1 (day~4816) STIS G750M
spectroscopy, which suggests that Spot~1 is now moderately resolved 
in the \hst\ data at optical wavelengths.

\subsection{Low Resolution Optical Observations} \label{opt-low}

\begin{figure}
\plotone{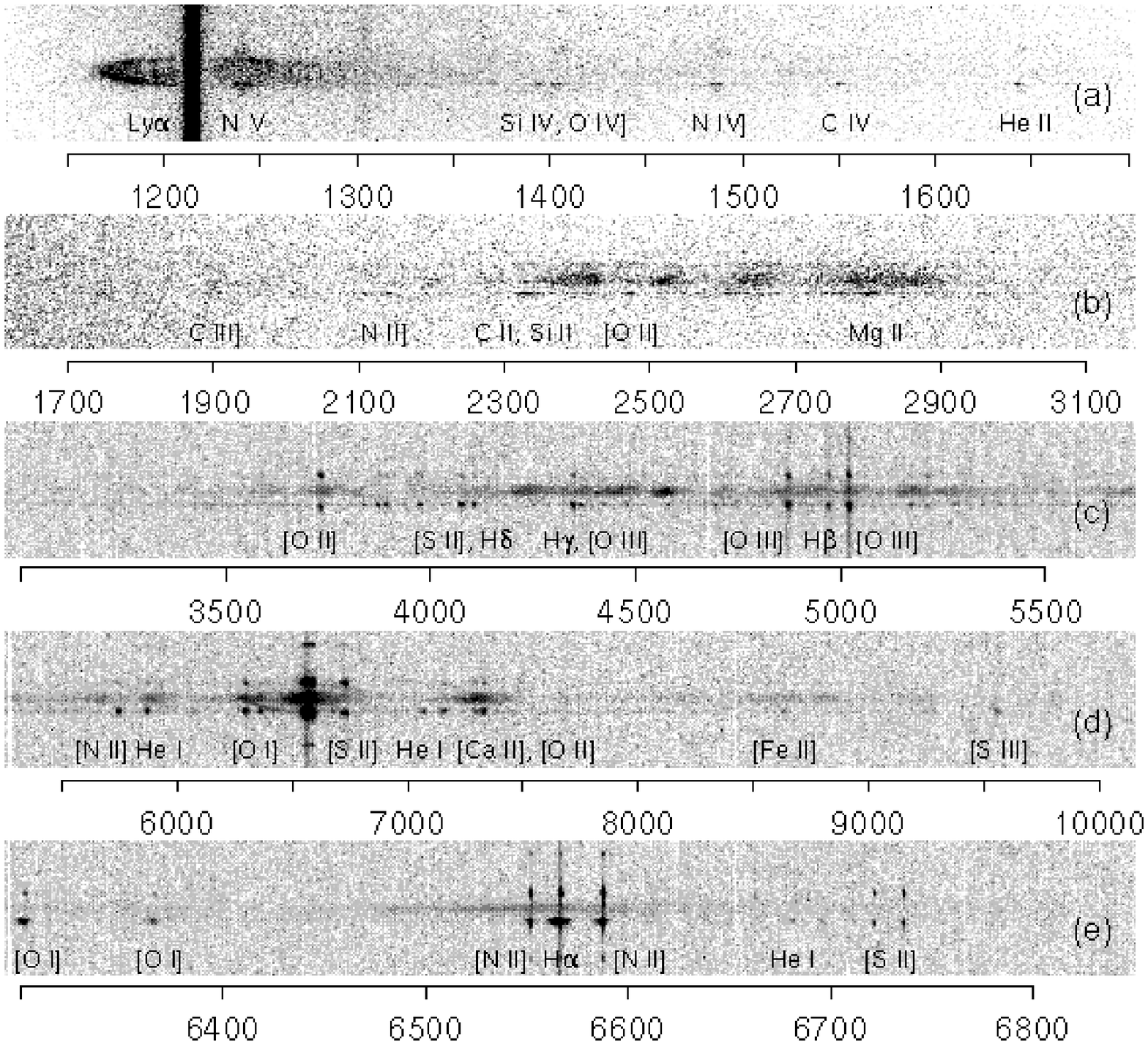}
\caption{
(a) Subsection of the 1999 October 7 STIS G140L (1425) 
spectral image showing the emission from Spot~1.
The height of the image displayed is 6\arcsec\ along the slit.
(b) Same as (a), for the 1999 September 17 STIS G230L (2376) data.
(c) Same as (a), for the 1999 September 3 STIS G430L (4300) data.
(d) Same as (a), for the 1999 September 18 STIS G750L (7751) data.
(e) Same as (a), for the 1999 August 30 STIS G750M (6581) data.
}
\label{fg-data}
\end{figure}

We extracted the low resolution optical spectrum of Spot~1 from the
STIS G430L and G750L two-dimensional spectral images. Portions of the
1999 September G430L and G750L data taken with the
0\farcs2 slit are shown in Figures~\ref{fg-data}(c) and
\ref{fg-data}(d), respectively.  The horizontal streaks near the center
in the figure are broad emission lines from the inner supernova
debris, which have a FWHM velocity $v_{\rm FWHM} \approx 2800 \kms$
\citep{wang96,chug97}.  For these data, the lower section of the
spectral image is the combined Spot~1 and inner ring emission-line
spectrum (hereinafter referred to as the Spot~1+North-Ring, or {\it
S1+NR\/}, spectrum).  At the kinematic resolution of the G430L and G750L 
grating settings ($\Delta V \approx 300 - 550 \kms$), neither the emission 
lines from Spot~1 [$v_{\rm FWHM} \approx 200 \kms$, \citet{mich00}] or those
from the ring ($v_{\rm FWHM} \simeq 10 \kms$) are resolved.  The upper
section of the spectral image is the emission-line spectrum of the
segment of the inner ring subsection that is in the slit and directly
opposite that of Spot~1 (hereinafter referred to as the South-Ring, or
{\it SR\/}, spectrum).

We measured the {\it S1+NR\/} and {\it SR\/} spectra in the G430L and
G750L grating settings by integrating the $7-9$ rows of the image in
which the emission-line spectra appeared. Emissions due to the diffuse
LMC background in the Balmer lines, [\oii] \wll3727, 3729, and [\oiii]
\wll4959, 5007 are observed as images of the entire slit at those
wavelengths in Figures~\ref{fg-data}(c) and \ref{fg-data}(d).
We subtracted the contribution of this diffuse emission
from the Spot~1 spectrum by linear interpolation above and below
the extracted rows.  Since the subtracted LMC background was only a
small fraction of the emission from Spot~1 ($<$2\% for [\oiii]
\wl5007), this subtraction did not significantly increase the
uncertainty of the estimated line fluxes.

We measured the flux of each emission line in the {\it S1+NR\/} and
{\it SR\/} spectra by fitting a Gaussian to the line profile, allowing
the net flux, width, and central wavelength to vary independently for
each line.  We estimated the background level from a linear fit over a
45-pixel region of the spectrum centered on the line in question but
excluding any emission-line features.  We fitted the line flux by
minimizing the total $\chi^{2}$ in which the coefficients defining the
line and the background were free parameters.  Gaussian profiles gave
satisfactory fits to all the line profiles.  The reduced-$\chi^2$, 
or $\chi^2_r$, (the total $\chi^2$ divided by the number of degrees 
of freedom), of our line fits were within the range $0.8-2.8$, 
compared to the value of 1.0 for a statistically good fit.  
We computed the flux of each line and its
error from the best-fit parameters and their associated uncertainties.
We adjusted the statistical errors of all the line fluxes so that 
$\chi^2_r = 1.0$ for all fits.  In cases where
two or more emission lines overlapped in wavelength, such as [\oiii]
\wl5007 + \hei\ \wl5016, and [\ion{Ar}{3}] \wl7136 + [\ion{Fe}{2}]
\wll7155, 7172, we fitted the emission features with multiple Gaussians
with the additional constraint that their wavelength separations were
the known differences of laboratory wavelengths.

Only a subset ($\sim 1/3$) of the emission lines observed in the {\it
S1+NR\/} spectrum also appeared in the {\it SR\/} spectrum.
Therefore, we attributed entirely to Spot~1 the measured fluxes of
emission lines that were seen in the {\it S1+NR\/} spectrum but not in
the {\it SR\/} one.  To subtract the {\it NR\/} spectrum from the {\it
S1+NR\/} spectrum, we assumed that the fluxes of emission lines in the
{\it NR\/} spectrum are equal to those in {\it SR\/} spectrum scaled
by factors that are independent of time.  This assumption is
reasonable because the rate of flux decrease around the ring has been
shown to be relatively constant around the ring \citep{plai95,lund01}.
We estimated the scale factors by examining archival WFPC2 emission-line
images in \Ha, [\oiii] \wl5007, and [\nii] \wl6583 obtained in 1994
February and 1994 September, before Spot~1 appeared.  We measured flux
ratios $f(NR)/f(SR)$ of 1.2, 1.1, and 1.3, respectively from the \Ha,
[\oiii] \wl5007, and [\nii] \wl6583 images.  We used the \Ha\ scale
factor for the Balmer lines.
While we used the [\oiii] \wl5007 factor for the
[\oiii] \wll4363, 4959, 5007 lines, and likewise for the [\nii]
lines, we recognized the increase in systematic errors in the
measured Spot~1 fluxes of [\oiii] \wl4363 and [\nii] \wl5755 because
these lines are more temperature sensitive than the other lines.
For all remaining emission lines, such as [\sii] and
[\ion{Ne}{3}], we assumed a scale factor of $1.2 \pm 0.1$ in order to
subtract the {\it NR\/} spectrum.

In addition to the 0\farcs2 data sets described above, we obtained one
observation of Spot~1 with the 0\farcs5 slit and the G750L grating
setting.  As before, we extracted the combined {\it S1+NR} spectrum by
integrating the 6 rows of the image where the emission lines appeared.
For the emission lines produced predominantly by Spot~1,
such as \hei\ \wl6678 and [\ion{Ar}{3}] \wl7135, we measured
the {\it S1} fluxes by fitting the line emissions with
single Gaussian profiles.  For the lines where
emission from both {\it S1+NR} were apparent, such as [\nii] \wl5755
and [\oi] \wl6300, we estimated the contribution of the {\it NR} flux
to the {\it S1+NR} flux from a linear interpolation of the {\it NR}
emission-line flux adjacent to Spot~1 in the slit.  After subtracting
this background, we fitted the remaining {\it S1} emission lines with
Gaussian profiles.  The uncertainties in the background ring flux
estimated in this procedure resulted in larger systematic errors in
the estimated Spot~1 line fluxes for this observation.

\subsection{Medium Resolution Optical Observations} \label{opt-med}

A medium resolution optical spectrum was taken on 1999 August 30 (UT,
4368.0 days since explosion) with a 0\farcs1 slit and the G750M
(6581) grating setting ($6295 - 6867$~\AA).  With a spectral
resolution of $\Delta V \simeq 50 \kms$, emission lines from Spot~1,
with $v_{\rm FWHM} \approx 200 \kms$ \citep{mich00}, were resolved in the
data, while the emission from the unshocked inner ring, with $v_{\rm
FWHM} \simeq 10 \kms$, remained unresolved.  Three observations of 7800
seconds each were taken at three parallel slit positions, pointed so
that the middle slit position was centered on Spot~1 and the two
other slit positions were immediately adjacent
[cf.\ Figure~\ref{fg-slitpos}(d)].  Therefore the
observation covered a segment of the ring of length 0\farcs3.

With the crowded stellar field near \sn, we did not execute the
acquisition-peakup exposure for these 0\farcs1 slit observations
as suggested by the \hst/STIS operation manual \citep{leit00}.
Instead all three adjacent slit positions were placed on Spot~1
by blind offsets.
Therefore we cannot apply the standard pipeline data reduction
procedures to process these data.
To determine the fluxes of emission lines from Spot~1, we first
removed the wavelength-dependent aperture throughput correction
function in the pipeline data for each 0\farcs1 slit spectrum.
We then summed the fluxes from the three 0\farcs1 slit
positions, and multiplied the total flux by a new aperture correction
function for an equivalent 0\farcs3 slit, calculated by interpolating
the pipeline corrections for the 0\farcs1, 0\farcs2, and 0\farcs5
slits.  For \Ha, the aperture correction led to a $\simeq20\%$
decrease of flux over the sum of the fluxes measured in the three slit
positions.  We calculated the corresponding 1$\sigma$ errors by
combining the individual errors in quadrature.

A section of the spectral image from the middle slit position is shown
in Figure~\ref{fg-data}(e).  Again, the central horizontal streak is
emission from the \sn\ debris.  Emissions from the inner ring at the
two positions where the ring intersected with the slit aperture were
observed in [\oi] \wll6300, 6364, \Ha, [\nii] \wll6548, 6583, and
[\sii] \wll6717, 6731.  In the lower section of the spectral image,
emission from the stationary inner ring overlapped with the broadened
emission from Spot~1.
Again, the major source of contamination the Spot~1 spectrum is
the emission from the inner circumstellar ring within the 0\farcs1 slit.
As described by \citet{mich00}, we fit all the Spot~1 emission
features with Gaussian functions.  Emission from the stationary
ring dominated the spectral profile near zero velocity
($\sim \pm 50\kms$) and was excluded from the fit.
The majority of the line profiles could be
fitted well with Gaussian profiles, such as the fit to the
[\nii]~\wl6583 line emission shown in Figure~\ref{fg-gaussfit}.
The sole exception was the \Ha\ line profile, where the signal was strong
enough to show noticeable departures from a Gaussian profile, as we
shall discuss further in \S\ref{sec-lineprof}.

\subsection{Low Resolution UV Observations} \label{uv}

We obtained low resolution UV spectra of Spot~1 with the
G140L and G230L grating settings.  \citet{mich00} have already
presented results from the first G140L far-UV observations in 1997
September 27 taken with the 0\farcs5 slit.  We detected emission lines
from Spot~1 in \nv\ \wll1239, 1243, \SiIV\ \wll1394, 1403, \oiv]
\wl1400, \niv] \wll1483, 1487, \civ\ \wll1548, 1551, and \heii\
\wl1640.  We detected the same set of emission lines in 1999 February
27, observing again with the 0\farcs5 slit.  In 1999 October 7,
observing with the 0\farcs2 slit, we also detected the \cii\ \wl1335
multiplet, [\ion{Ne}{4}] \wl1602, and \oiii] \wll1661, 1666.
Figure~\ref{fg-data}(a) shows a section of the spectral image from
this observation.  Radiation from Spot~1 is evident in the lower
portion of the image whereas only faint line emission from the inner
circumstellar ring can be seen in the upper half of the displayed
image.  The broad ($\sim \pm 15,000 \kms$) \Lya\ radiation comes from
the reverse shock from the interaction between the supernova debris
and the \ion{H}{2} region located inside of the equatorial ring
\citep{sonn98,mich98a,mich98b}. Fluxes of UV line emission from the 
inner ring are much less than those from Spot~1 and make a negligible
contribution to the measured fluxes.  This is
also the case for the near-UV emission lines measured in the 1999
September 17 G230L 0\farcs2 observation, shown in
Figure~\ref{fg-data}(b).

We measured the far-UV and near-UV spectra of Spot~1 from the G140L and
G230L data, respectively, by procedures similar to those we described in
\S\ref{opt-low}.  We fitted emission lines with Gaussian profiles and
linear backgrounds, except for the broad \Lya\ emission underlying the
\nv\ \wll1239, 1243 doublet which we fitted with a quadratic function.
We fitted the two components of close doublets such as \nv\ \wll1239,
1243, \civ\ \wll1548, 1551, \oiii] \wll1661, 1666, \nii] \wll2139,
2143, and \mgii\ \wll2796, 2803 with Gaussians constrained to have
fixed doublet separations, identical widths, and line ratios dictated
by their oscillator strengths.  At the resolution of the G140L grating
setting, the \SiIV\ \wl1403 emission of the \SiIV\ \wll1394, 1403
doublet is blended with the \oiv] \wl1400 multiplets.  We fit the
combined \SiIV\ and \oiv] feature with multiple Gaussians, requiring
the \SiIV\ doublet to meet the same constraints as the other close
doublets.

The observed fluxes of a few UV lines, such as \SiIV\ \wll 1394, 1403
and \civ\ \wll1548, 1551, were reduced by interstellar line
absorption.  We describe our procedure for correcting for this
absorption in \S\ref{uv-lines} below.

\section{Data} \label{sec-data}

\subsection{Optical Emission Line Fluxes} \label{opt-flux}

Table~\ref{tb-optflux} lists the measured Spot~1 optical emission line
fluxes, including previously published results from the 1998 March 7
(day 4030) data by \citet{mich00} and 1$\sigma$ upper flux limits for
the [\oii] \wll 3726, 3729, and [\ion{N}{1}] \wll 5198, 5200 doublets.
The 1$\sigma$ errors tabulated are only statistical errors.
Systematic effects, such as fringing for the G750L data towards the
near-IR region ($\lambda > 8500$~\AA, Leitherer et al.\ 2000), might
contribute additional uncertainties to the measured fluxes.

The tabulated fluxes have been dereddened with E($B-V$) of 0.16
\citep{fitz90} and the extinction law of \citet{card89}
with an assumed R$_{V}$ of 3.1.  In the optical band the
differences between the LMC extinction law and the Galactic law are
negligible at low color excess \citep{fitz99}.  The interstellar
extinction correction applied is listed in the last column of
Table~\ref{tb-optflux}.

Spot~1 was observed in many neutral and lowly ionized species in the
optical wavelengths.  We did not detect any coronal lines such as
[\ion{Fe}{10}] \wl6375.  Most of the line fluxes increased with time
at a rate comparable to that measured from WFPC2 photometry
\citep{garn01}. At the low spectral resolution of both the G430L and
G750L observing modes ($R = \lambda/\Delta\lambda = 530 - 1040$),
definitive line identification remained a problem, especially for the
[\feii] emission lines.  The Fe line identifications in
Table~\ref{tb-optflux} are based upon the modeling of the Spot~1 Fe
lines in Pun et al. (2002, in preparation).  Moreover, several
lines were blended.  Table~\ref{tb-optflux} lists possible
contributing species and, in the case of [\feii] lines, different
multiplets.  In contrast, line blending is not a problem in the medium
resolution G750M ($R \approx 6000$) data.

Fluxes in [\oi] \wll6300, 6363 and \hei\ \wl6678 were measured with
the low resolution G750L and medium resolution G750M gratings in 1999
September within 17 days of each other.  The two measurements agreed
within uncertainties for the [\oi] doublet. The two results
differed by $\sim 50\%$ for the low S/N \hei\ \wl6678 data,
but were also within the noise level.
This difference is probably a good indication of the
detection limits of such faint lines.

\subsection{Optical Emission Line Widths} \label{opt-width}

For the medium resolution G750M observations, apart from the line 
fluxes, we were also able to measure the
peak emission velocities ($V_{0}$) and the widths ($V_{\rm FWHM}$) of
the emission lines from the profile fits.  Table~\ref{tb-optwidth}
lists the results.  The peak emission velocity measurements 
have been adjusted for the \sn\ heliocentric velocity of $+286\kms$ 
\citep{wamp89}. 
The 1998 March 0\farcs2 G750M results have also been adjusted for 
the off-center position of Spot~1 within the aperture (cf. \S\ref{sec-obs}).
The main
uncertainties in the measurements of both $V_{0}$ and $V_{\rm FWHM}$
are due to the contributions by emission from the stationary
circumstellar ring, which dominated the emission by Spot~1 near zero
velocity for many species (cf. Figure~\ref{fg-gaussfit}).  The errors
due to this contribution are generally smaller for the 1999 August
0\farcs1 observations than the 1998 March 0\farcs2 ones.

For all emission lines, the line centroids from Spot~1 were
blueshifted, with peak velocity $V_0$ lying within the range $-40$ to
$-10\kms$.  This result is consistent with the overall physical
picture in which Spot~1 is located on the near side of the equatorial
ring \citep{sonn98} and the shock entering Spot~1 is moving towards
the observer.  We found no evidence that the peak velocity $V_0$ of
Spot~1 varied with time.

Most emission lines had widths (FWHM) within the range $\sim 150 - 180
\kms$, except [\nii] \wl6583 and \Ha, which had a slightly greater
width, $V_{\rm FWHM} \gae 200 \kms$.  
We detected no measurable change with time of
the line widths except for \Ha\ and [\nii] \wl6548.  For \Ha, the
emission profile from the second observation in 1999 August could no
longer be fit well by a single Gaussian (to be discussed further in
\S\ref{sec-lineprof}).  The width of the [\nii] \wl6548 emission,
measured in 1999 August was 1.5~times greater than that measured in
1998.  However, we are inclined to attribute this increase to
systematic error, since we detected no such increase in the other
[\nii] component at 6583~\AA, where the line fluxes were $\sim 3$
times stronger.

\subsection{UV Emission Lines} \label{uv-lines}

\subsubsection{Interstellar Line Absorption} \label{line-abs}

Near the time of outburst, interstellar absorption lines of \cii\
\wl1335 multiplet, \SiIV\ \wll1394, 1403, \civ\ \wll1548, 1551, and
\mgii\ \wll2796, 2803 were observed in the UV continuum of
\sn\ with \iue\ operating in the high resolution ($\Delta V \approx
30\kms$) echelle mode \citep{blad88,welt99}.  For each line, the
dominant absorption component was centered at $+281\kms$, near the
\sn\ heliocentric velocity of $+286\kms$ \citep{wamp89}, and had a
FWHM of $\sim 80\kms$.  Therefore, narrow emission lines ($v_{\rm FWHM}
\simeq 10\kms$) from the inner circumstellar ring from these species
are completely blocked by the interstellar absorption \citep{fran89},
as demonstrated by the absence of \SiIV\ and \civ\ emission lines from
the ring in Figure~\ref{fg-data}(a).

\begin{figure}
\plotone{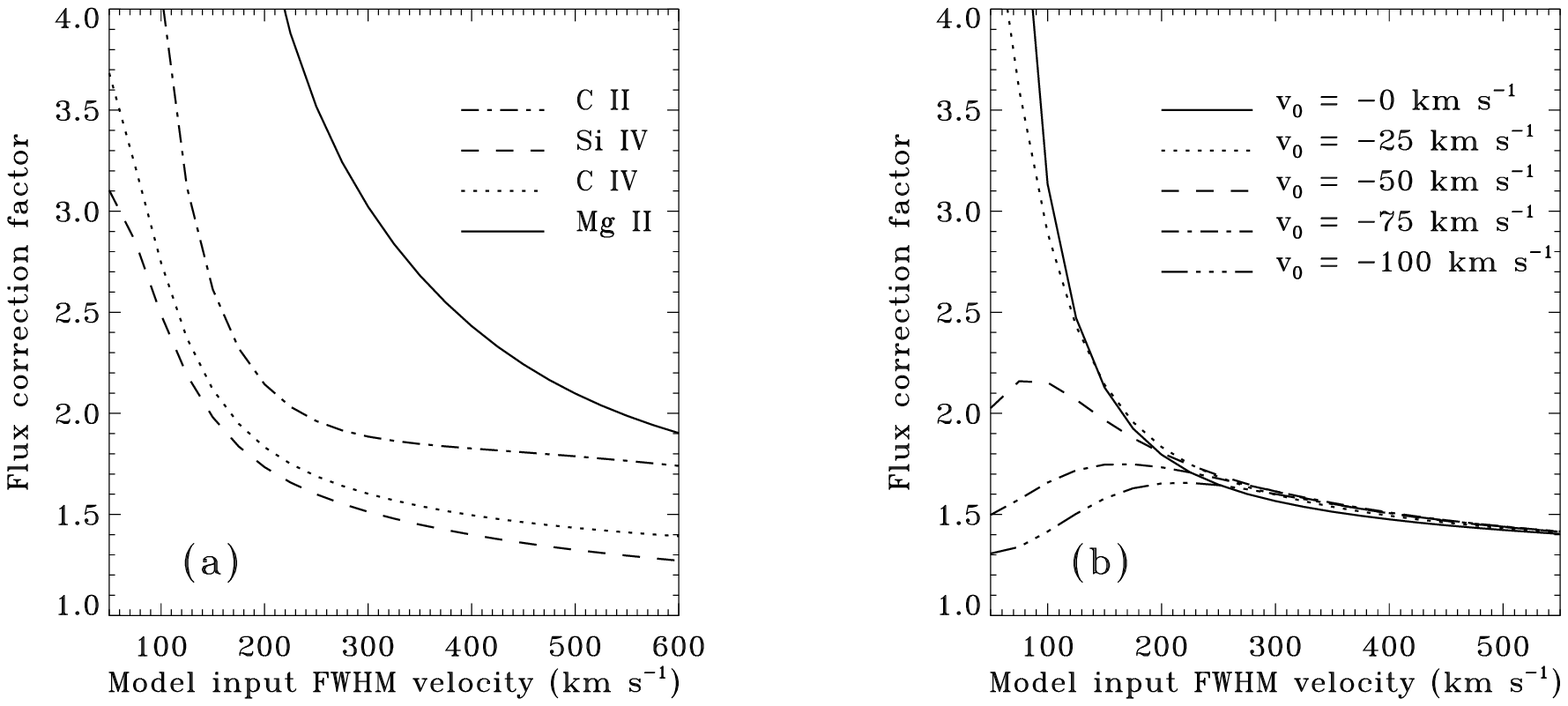}
\caption{
(a) The interstellar absorption correction factors for emission lines
\cii\ \wll1335,1336 ({\it dot-dash}), \SiIV\ \wll1394, 1403
({\it dash}), \civ\ \wll1548, 1551 ({\it dot}), and \mgii\ \wll2796, 2803
({\it solid}).
The peak velocity, $V_0$, of the model Spot~1 emission is assumed to be
$-30\kms$.
(b) The correction factor for \civ\ \wll1548, 1551 doublet is
plotted against the FWHM velocity of the input model Spot~1 Gaussian
profile with peak velocities $V_{0} = 0\kms$ ({\it solid}),
$-25\kms$ ({\it dots}), $-50\kms$ ({\it dash}),
$-75\kms$ ({\it dot-dash}), and $-100\kms$ ({\it dot-dot-dot-dash}).
}
\label{fg-fcorr}
\end{figure}

Emission lines from Spot~1, with $V_{\rm FWHM} \approx 200\kms$, 
are not totally blocked by these interstellar absorption lines, 
as correctly predicted by \citet{luo94}.  
However, the line profiles are altered and the
observed fluxes are reduced substantially.  Therefore, we must correct
the measured \cii, \SiIV, \civ, and \mgii\ emission line fluxes to
account for this absorption.  The appropriate correction factors
depend on the profile shapes of both the Spot~1 emission and the
intervening absorption.  To make this correction, we assumed that the
UV emission lines from Spot~1 had Gaussian profiles with the same
parameters as the optical lines as measured in the medium resolution
optical data (\S\ref{opt-med}).  
We then estimated the amount of flux reduction in the UV emission 
lines by multiplying the assumed Gaussian profiles by
the absorption profiles seen in the \iue\ data.  The corresponding
flux correction factors for the various emission lines are shown in
Figure~\ref{fg-fcorr}(a).  The correction factor is greatest for the
\mgii\ \wll2796, 2803 doublet, where the interstellar absorption is
almost completely saturated between $-50$ and $+300 \kms$.

The flux correction factor due to interstellar absorption is sensitive
to the assumed widths ($V_{\rm FWHM}$) and peak ($V_{0}$) velocities
of assumed line profiles from Spot~1. Figure~\ref{fg-fcorr}(b)
illustrates this dependence for the important case of \civ\ \wll1548, 1551.  
We see that the flux correction factor increases
moderately with decreasing $V_{\rm FWHM}$ for $V_{\rm FWHM} >
200\kms$, but becomes very sensitive to both $V_{\rm FWHM}$ and
$V_{0}$ for $V_{\rm FWHM} < 150\kms$.

As we shall discuss below in \S\ref{sec-lineprof}, the peak velocities
and widths of the emission lines from Spot~1 depend on the detailed
geometry and hydrodynamics of the shocks entering the spot, which are
unknown.  It is not obvious that the optical and UV emission lines
should have the same peak velocities and widths.  However, in the
plane-parallel shock model that we describe in \S4, we found that the
peak velocities of emission from Spot~1 were almost identical in both
the UV and optical wavelengths.  Therefore, we used the measured peak
Spot~1 velocities from the medium resolution optical observations,
$V_0 = -30 \pm 15\kms$ (cf. Table~\ref{tb-optwidth}), to estimate the flux
correction factors to account for interstellar absorption of the UV
emission lines.

\subsubsection{Widths of the UV Emission Lines} \label{mcarlo}

The far-UV emission lines from Spot~1 are poorly resolved at the
spectral resolution of the G140L grating 
($\Delta V \approx 250 \kms$ at 1500~\AA).  
For these lines, we attempted to establish the relation
between the actual and observed widths through Monte Carlo
simulations.  For the simulations, we assumed that the intrinsic
Spot~1 emission lines had (i) a Gaussian shape with FWHM as a
variable parameter, (ii) emission peaked at $-30 \kms$, and (iii) flux
ratio of the doublets dictated by their oscillator strengths.

\begin{figure}
\plotone{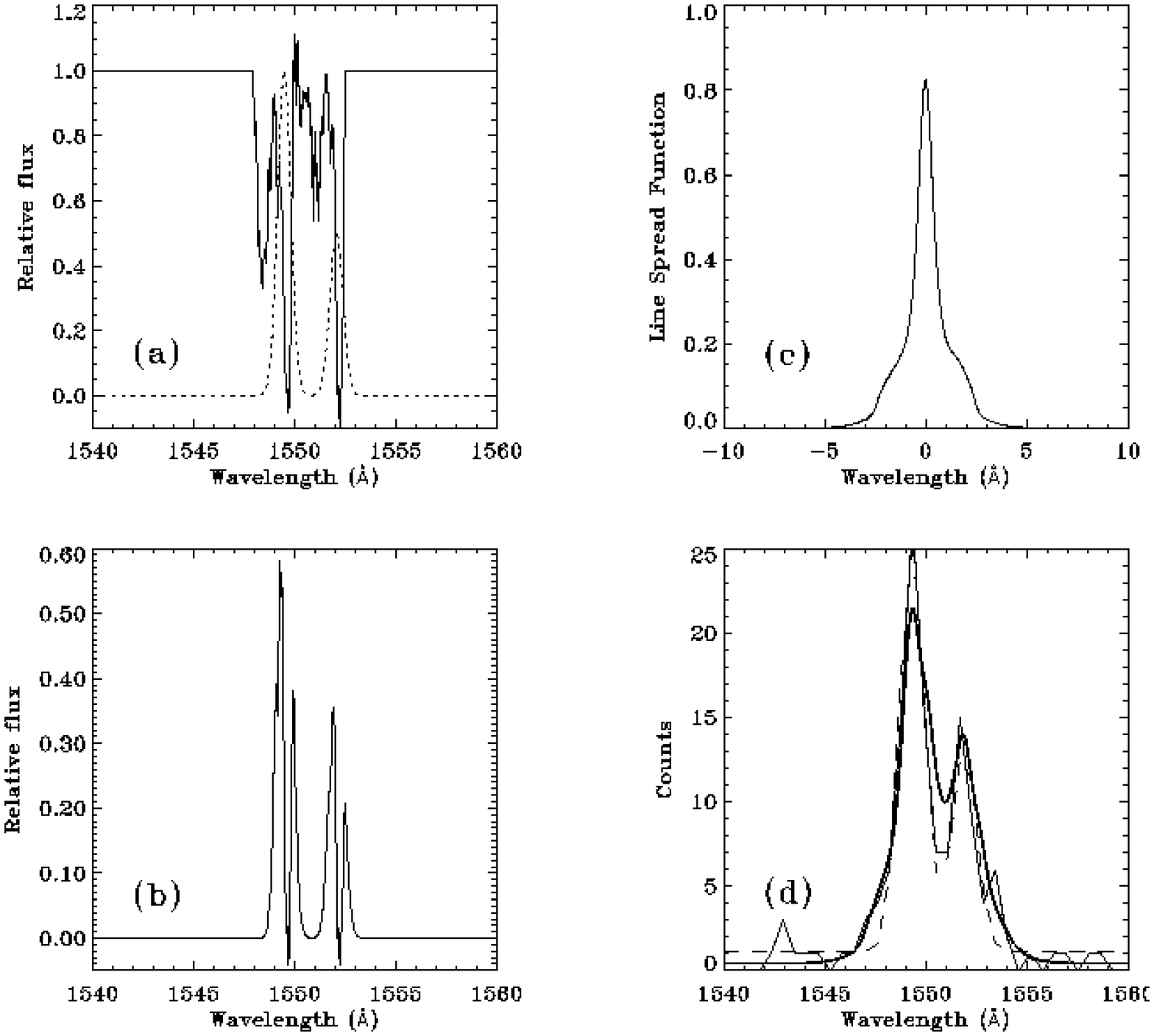}
\caption{
(a) The \iue\ high dispersion spectrum
of the \civ\ \wll 1548, 1551 absorption lines ({\it solid}), overlayed
with a model Gaussian Spot~1 line profile of the \civ\ emission line
({\it dots}). The FWHM width of the model profile is $150\kms$ and it peaks
at $-30\kms$ with line ratio $I$(1548)/$I$(1551) = 2:1.
(b) The model \civ\ Spot~1 spectrum after the interstellar line absorption.
(c) The LSF of STIS/MAMA G140L grating for the 0\farcs2
aperture at 1500~\AA.
(d) The model ``observed'' spectrum after convolving the model profile
after absorption with the line spread function ({\it thick}). Example of a
Monte-Carlo generated line profile of the \civ\ doublet, normalized
to the integrated counts of the 1999 October data, is also shown ({\it solid}),
along with the best-fit double Gaussian function to this profile 
({\it dash}). In this example, the best-fit $V_{\rm FWHM} = 281\kms$.
}
\label{fg-absorb}
\end{figure}

The dotted curve in Figure~\ref{fg-absorb}(a) shows a model \civ\ \wll
1548, 1551 profile, assuming $V_{\rm FWHM} = 150 \kms$ and line flux ratio
$I(1548)/I(1551) =$ 2:1.  The solid curve shows a subsection of the
high resolution \iue\ spectrum with the absorption line profile for
the \civ\ doublet.  For each input FWHM, we multiplied the model input
emission line by the measured \iue\ absorption spectrum.
Figure~\ref{fg-absorb}(b) shows a typical result.  We convolved this
profile with the line-spread function (LSF) of the detector for the
slit aperture of the data set.  The LSF for each emission line was
interpolated from the measured LSFs at 1200~\AA, 1500~\AA, and
1600~\AA. The LSF typically has a narrow peak ($\sim 1.5$ pixel) and a
broad wing that extends to $\sim 10$ pixels on either side of the peak
\citep{leit00}.  Figure~\ref{fg-absorb}(c) shows a typical example of
the LSF for the G140L grating at 1550~\AA\ with the 0\farcs2 aperture.

For each emission line, we normalized the resulting convolved profile
to the photon counts for each observation.  The thick solid curve in
Figure~\ref{fg-absorb}(d) shows the normalized model \civ\ line
profile for the 1999 October observation.  We constructed a simulated
observed profile (the thin solid curve in Figure~\ref{fg-absorb}d)
by applying random noise to the model profile and sampling it at the
spectral resolution of the G140L grating.  Finally, we fitted the
simulated profile with Gaussians in the same way as the real data.
The dotted curve in Figure~\ref{fg-absorb}(d) shows such a fit.

\begin{figure}
\plotone{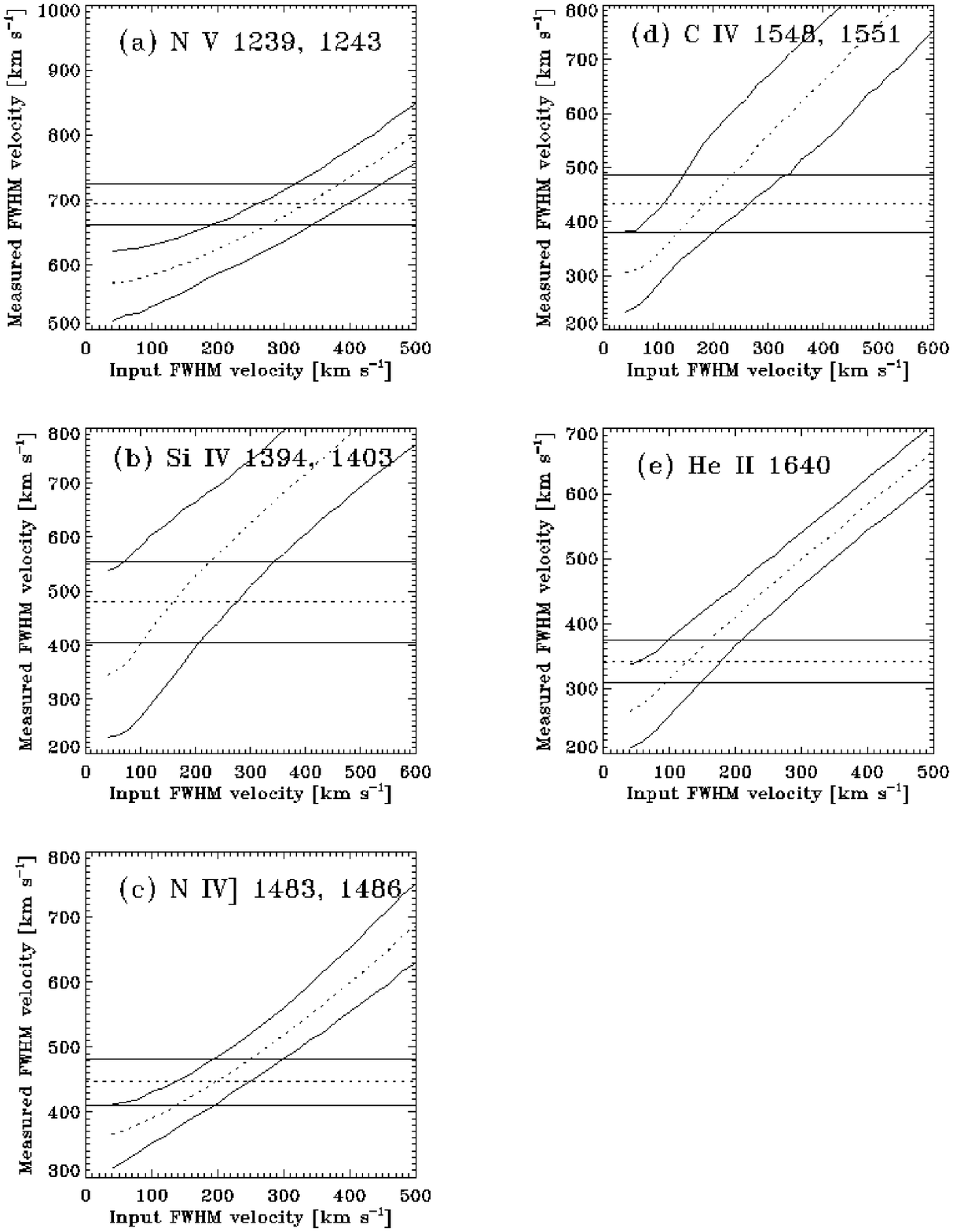}
\caption{
Monte Carlo results of the relation between the intrinsic Spot~1 emission
width and the measured widths from Gaussian fittings for the 1999 October
G140L 0\farcs2 observations in
(a) \nv\ \wll1239, 1243;
(b) \SiIV\ \wll1394, 1403; (c) \niv] \wll1483, 1486;
(d) \civ\ \wll1548, 1551; and (e) \heii\ \wl1640.
}
\label{fg-mc}
\end{figure}

For each input model FWHM velocity, we ran 10000 simulations
and generated an array of the corresponding observed line widths.
Figure~\ref{fg-mc} shows the median and the 68.3\% (1$\sigma$) upper
and lower limits of the array plotted against the model input widths
of the various observed emission lines.  We use these results to
convert the observed line widths from the STIS data, shown as
horizontal lines in Figure~\ref{fg-mc}, to the intrinsic widths of the
lines emitted by Spot~1.

We did not attempt to model the near-UV G230L data in this way because
the spectral resolution for this grating setting ($\Delta V \approx 450
\kms$ at 2500~\AA) was too low for such a procedure to yield useful
results.

For all emission lines except the \nv\ doublet, we found a 1$\sigma$
upper limit of ${\rm FWHM} \lae 300 \kms$, consistent with the
measurements from the medium resolution optical data.  The lower limit
to the assumed FWHM is important because the flux correction factors
for all UV emission lines are sensitive to the input line widths for
FWHM $ < 150 \kms$.  We took this lower limit to be the same as the
measured lower limit for the optical lines, that is, ${\rm FWHM}> 100
\kms$.

\subsubsection{UV Line Fluxes and Line Widths} \label{uv-flux}

Table~\ref{tb-uvflux} lists the fluxes of UV lines from Spot~1
inferred from the Gaussian line fits (corrected for extinction) and
the intrinsic line widths derived from the Monte Carlo simulations.
The tabulated fluxes for the \cii, \SiIV, \civ, and \mgii\ doublet
have been corrected for interstellar line absorption assuming an
intrinsic Spot~1 line width $V_{\rm FWHM} = 150^{+150}_{-50} \kms$ with
peak emission at $V_0 = -30 \pm 15 \kms$.  The flux correction factors
applied for these lines are also listed in Table~\ref{tb-uvflux}.
We also corrected the fluxes for interstellar extinction by assuming
$E(B-V)_{\rm LMC} = 0.06$, $E(B-V)_{\rm Galactic} = 0.10$, and $R_V = 3.1$.
At UV wavelengths, the choice of extinction function is
important because the correction is substantial and is known to vary
from place to place (cf. Pun et al.\ 1995).
We used the 30 Doradus extinction function of \citet{fitz86}
for the LMC component,
and the \citet{seat79} function for the Galactic component.

The far-UV G140L data had slightly higher kinematic resolution ($\Delta
V \approx 250 \kms$) than the low resolution optical G430L and G750L
data.  However, the individual components of \cii\ \wl1335 and
\oiv]\wl1400 multiplets remained unresolved in the data.  There were more
uncertainties with line identifications in the lower resolution
($\Delta V \approx 300 - 650 \kms$) near-UV G230L data, such as the
unidentified emission features near 2737~\AA\ and 2746~\AA.  On the
other hand, we identified emission features near 2324~\AA\ and
2334~\AA\ as the \cii\ \wl2325 multiplet and \SiII\ \wl2335,
respectively.  We ruled out the alternative identification of [\oiii]
\wll2322, 2332 because the observed line ratios, 
$I(2324$~\AA)/$I(2334$~\AA) $\approx 5$ and 
$I(2324$~\AA)/$I(4363$~\AA) $\approx 7$, 
were much different from the theoretical [\oiii] line ratios of
$\sim 280$ and $0.12$, respectively. These [\oiii] line ratios are
determined only by the atomic transition probabilities and are
independent of the excitation model.

The fluxes of all far-UV lines from Spot~1 increased with time during
the three observations taken from 1997 September (day 3869) to 1999
October (day 4606).  The rate of increase differed for features from
different ions, ranging from $I$(3869~d)/$I$(4596~d) $= 5.3$ for
\nv\ \wl1240 to 1.9 for \civ\ \wl1550.  We will discuss the
time dependence of the UV emission-line fluxes in \S\ref{sec-shockmodels}.

\section{Interpretation}  \label{sec-interp}

\subsection{Impact Hydrodynamics} \label{sec-globalhydro}

In our working model for Spot~1, we assume that the UV and optical
emission lines observed are caused by radiative shocks that
develop where the supernova blast wave strikes dense gas protruding 
inward from the circumstellar ring.
As we shall show, the spectrum and profiles of the emission lines from
such shocks are sensitive to the density distribution and geometry
of this protrusion, which are probably quite complex and cannot be
resolved even with the \hst.
Our approach here therefore is to illustrate the salient physics
of the spectrum formation with a few idealized ``toy models," which
we believe will guide us toward a better understanding of
the properties of a more realistic model.

Following the previous work of \citet{luo94} and \citet{bork97b},
we show in Figure~\ref{fig:globalhydro} hydrodynamic simulations
for two models of a fast shock overtaking a dense gas cloud.
In each model, we assume that a fast plane-parallel blast wave
traveling through a uniform medium of relatively low density
overtakes a cloud of substantially greater uniform density.
The cloud boundary is approximated as a density discontinuity.
In each simulation, the blast wave drives a transmitted shock
into the cloud, while a reflected shock travels backwards
towards the interior of the remnant.
As the blast wave overtakes the obstacle, the surface area
of the transmitted shock increases.  The transmitted shock
propagates with a range of velocities depending on shape and
density of the obstacle.

The simulations are calculated using the hydrodynamic code VH-1
\citep{stri95}, which is based on the piecewise parabolic method 
of \citet{cole84}.
Radiative cooling is included in the code using an operator
splitting technique.
We have used a non-equilibrium ionization cooling
curve calculated with the plane-parallel shock code described in
\S\ref{sec-shockstruc} with abundances typical of the equatorial ring.
At the resolution of the simulations, we are not always able
to resolve the cooling time scale of the shocks.
However, this computational limitation does not seriously
affect the behavior of the overall hydrodynamics of the interaction.
Moreover, we have modeled the blast wave as a single planar shock
rather than using the full double-shock structure present in the
remnant.  We will comment on the effects of ignoring the full 
double-shock structure in \S\ref{sec-disc}.

The two scenarios depicted in Figure~\ref{fig:globalhydro} 
show two different behaviors for the development of radiative shocks 
in the obstacle.  The important parameter distinguishing their 
behavior is $t_{cool} / t_{cross}$, where $t_{cool}$ is the 
typical cooling time for the transmitted shocks, and $t_{cross}$ 
is the time it takes for the fast shock to cross the obstacle.  
As we will discuss below in \S\ref{sec-twoshock}, the interpretation
of the observed light curves of lines emitted from the shocks depends on
which behavior is dominant in Spot~1.

The simulation on the left of Figure~\ref{fig:globalhydro} (Scenario~1)
is one in which $t_{cool}$ is comparable to $t_{cross}$, so that not 
all of the shocks have become radiative.  
For this simulation we assume that the obstacle is a spherical cloud
of density $\rho_0 = 10^4 \amucc$ and diameter 
$d_{spot} = 4 \times 10^{16} \cm$.  This geometry allows the
the shock to be driven into the back side of the obstacle.
Even after the blast wave has completely overtaken the obstacle, the
shocks transmitted into the front-end of the obstacle have not
yet undergone thermal collapse and are non-radiative (NR)
because of the high impact velocity.
On the other hand, the shocks transmitted into the sides and
back of the obstacle have lower velocities owing to the oblique
incidence of the blast wave.
The shocked gas behind these parts of the
transmitted shock has a lower temperature and a shorter
radiative cooling time.
A dense layer of cooled gas (R) is developed as a consequence
(to be discussed further in \S\ref{sec-shockmodels}).

Scenario~2 in Figure~\ref{fig:globalhydro} represents
an example in which the cooling time for the transmitted
shock is much shorter than the time scale for
the blast wave to cross the cloud $t_{cool} \ll t_{cross}$, 
regardless of incidence angle of the blast wave. 
We assume that the obstacle is an
elongated protrusion with a higher density, $\rho_0 = 10^5 \amucc$.
With the high density of the obstacle, all the transmitted
shocks undergo thermal collapse soon after impact.

We recognize that these idealized models in
Figure~\ref{fig:globalhydro} do not represent the true shape
and density distribution of the obstacle.
However, given the limited observations at hand, we believe
it is more fruitful to explore how well we can fit the data
with a few idealized models or combinations thereof, rather
than to explore fits of more complicated
hydrodynamic models (see \S\ref{sec-disc}).

\subsubsection{Transmitted Shock Velocities} \label{sec-velrange}

The driving pressure $P_b$ behind a hypersonic blast wave propagating
with velocity $V_b$ through a medium of density $\rho$ is given by
$P_b = 3/4 \rho V_b^2$.  Immediately after the blast wave passes
the surface of an obstacle, a transmitted shock propagates into
the obstacle with velocity
\[
V_s = (P_s / \rho_0 )^{1/2} \ ,
\]
where $P_s$ is the pressure at the surface of the obstacle and
$\rho_0$ is the pre-shock density of the obstacle.
The value of $P_s$ at a given location on the cloud surface rises shapely
as the blast wave first strikes it, then decreases as the
reflected shock propagates away from it.
Over a time scale comparable to that for the blast wave to overtake
the obstacle, $P_s$ decreases by a factor of $\sim 1.5 - 2$
\citep{bork97b}.
Immediately after the passage of the blast wave, the ratio $P_s/P_b$ at
any given point on the cloud surface depends on two parameters,
first, the obliquity angle $\theta$ between the direction of the blast
and the inward normal to the surface of obstacle, and second,
the pre-shock density ratio $\delta = \rho_0/\rho_{HII}$ between 
the obstacle and the ambient \hii\ region.
Figure~\ref{fig:beta} illustrates the dependence on obliquity
for a blast wave with a density contrast $\delta \approx 70$ 
impacting on a spherical cloud.
At the point of first contact, i.e., $\theta = 0$, the driving
pressure is at its maximum value and $P_s/P_b \approx 4$.
As the obliquity of the impact increases, the driving pressure
decreases.
With a factor of $\sim 10$ decrease in pressure due to obliquity,
the initial velocity of the transmitted shock on the edges of the
protrusion is only $\sim30\%$ of that at the tip of the protrusion.

We estimate the pressure behind the blast wave, $P_b$, from the
relation $P_b = 3/4 \rho V_b^2$.
By fitting the radio remnant of SN~1987A from year 1992 to 1995
with a shell model, \citet{gaen97} measured the velocity of
the blast wave to be $V_b = 2800 \pm 400 \kms$.
Using more recent radio observations, \citet{manc01} updated
the value of $V_b$ to be $3500 \pm 100 \kms$.
However, they also noticed that in data taken since $\sim 1998$,
the radio remnant can no longer be fit well with the simple shell
model but instead with a combined shell and multiple point-sources model.
\citet{manc01} estimated that this leads to a 30\% uncertainty in 
the measured expansion velocity.
On the other hand, by modeling the observed X-ray emission from
the blast wave before any spots appeared on the circumstellar ring,
\citet{bork97b} found a slightly higher velocity for the
blast wave, $<V_b> = 4100 \kms$, and a density of the \hii\ region
inside the ring, $\rho_{HII} = 150 \amucc$.  These results are
consistent with the estimates of \citet{chev95} and the upper limit
$\rho_{HII} < 240 \amucc$ determined by \citet{lund99}.
We estimate the blast-wave pressure to be within
the range $P_b = (1$ -- $5) \times 10^{-5} \pres$.

The pre-shock density, $\rho_0$, of the obstacle is even
more uncertain.
The density of the ring has been determined from the rate
of fading of the optical and UV emission lines \citep{lund96,sonn97}.
At the time of emission maximum ($\sim$~day 350), the radiation
was dominated by gas of relatively high density,
$\rho_0 \sim 5 \times 10^4 \amucc$.
However, the emission from the higher density gas faded rapidly and
the emission at later times was dominated by gas of lower density,
$\rho_0 \sim 10^4 \amucc$ \citep{mara00}.
The gas in Spot~1 might have even higher density than the value
derived from optical and UV lines.
The fact that Spot~1 lies inside the inner circumstellar ring
might be due to the fact that it resisted ablation
from the ionizing radiation and stellar wind of the progenitor as a
result of its enhanced density.

We summarize the possible range of transmitted shock velocities in
Figure~\ref{fig:phasespace1}.
The shaded gray region shows the range of transmitted
shock velocity, $V_s$, as a function of the pre-shock density of the
obstacle, $\rho_0$, for our best estimate
of the blast-wave pressure ($V_b = 3500 \kms$,
$\rho_{HII} = 150 \amucc$).
The upper boundary is the velocity of the transmitted shock at
the tip of the protrusion and the lower boundary
is the velocity at the side.
We also show in Figure~\ref{fig:phasespace1}
the corresponding shock velocity range for our high estimate
of the blast-wave pressure ($V_b = 4100 \kms, \rho_{HII} = 250
\amucc$, {\it dashed}), and our low pressure estimate
($V_b = 2800 \kms, \rho_{HII} = 100 \amucc$, {\it dotted}).
Assuming a pre-shock obstacle density of $(1-5) \times 10^4 \amucc$,
it is apparent that a range of shock velocities
($\sim 100 - 1000 \kms$) can be present in the obstacle.

\subsection{Shock Structure} \label{sec-shockstruc}

In this section we review the relevant physics of shock fronts 
that we need to interpret the emission-line spectrum of Spot~1.  
We use the 1991 version of the Raymond shock code \citep{cox85} 
for illustration.
The code calculates the non-equilibrium ionization and excitation of
the post-shock flow in a one-dimensional steady state shock
for any given input parameters such as shock velocity and 
pre-shock densities.
For any specified electron ($T_e$) and ion ($T_i$) temperatures 
at the shock front, the code calculates their equilibration 
through Coulomb collisions. 
We assume a ratio of $T_e / T_i = 0.2$ for our models. 
The code calculates the upstream equilibrium preionization and the 
downstream photoionization of the gas and derives the post-shock density, 
temperature, and ionization structure in the gas.  
The code also calculates the local emissivity throughout the shock 
and the integrated fluxes over the entire column behind the shock
front, including continuum and line emission from both allowed and
forbidden transitions of H, He, C, N, O, Ne, Mg, Si, S, Ar, Ca, Fe, and Ni.

We adopted the LMC abundances measured by \citet{russ92} in
all elements except for He, C, N, and O, where the ring abundances
derived by \citet{lund96} are used instead. 
The other exception is Si where the abundance of \citet{welt99} 
is used because of the large uncertainty in the measurements by 
\citet{russ92}.  
We call this set of values the ``Ring'' abundance 
(H : He : C : N : O : Ne : Mg : Si : S
: Ar : Ca : Fe : Ni = 1 : 0.25 : 3.24 $\times 10^{-5}$ : 1.82 $\times
10^{-4}$ : 1.58 $\times 10^{-4}$ : 4.07 $\times 10^{-5}$ : 2.95
$\times 10^{-5}$ : 2.51 $\times 10^{-5}$ : 5.01 $\times 10^{-6}$ :
1.95$\times 10^{-6}$ : 7.76 $\times 10^{-7}$ : 1.70 $\times 10^{-5}$ :
1.10 $\times 10^{-6}$).

We show the post-shock temperature and density structure for 
protons and electrons in our model shock in the upper panel of
Figure~\ref{fig:shock}. 
We assume a shock velocity of $V_s = 250 \kms$, characteristic 
of the shocks producing the radiation seen from Spot~1 \citep{mich00}. 
The shock has a high Mach number ($M = V_s / c_s \gg 1$), 
and compresses the obstacle gas by a factor of $\sim$~4 at
the shock front, providing that the magnetic field is negligible.  
The temperature of the ions after crossing the shock front is given as
$T_i = 3 m_i V_s^2 / 16 k$, where $m_i$ is the mass of the ion and
$k$ is the Boltzmann's constant.  
This implies that each ionic species will have a different 
post-shock temperature, provided that there are processes to
completely thermalize the post-shock distribution functions. 
For collisional shocks, the post-shock electron temperature 
is orders of magnitude lower than the ion temperatures \citep{zeld67}. 
On the other hand, in collisionless shocks, plasma turbulence 
(e.g., \citealt{carg88}) can partially equilibrate the post-shock 
electron and ion temperatures. 
Eventually, the temperatures will be equilibrated by
Coulomb collisions to a mean shock temperature 
$T_s = 3 \mu m_p V_s^2/ 16 k$, where $\mu = 4/(8-5Y)$ is the mean 
atomic weight per particle and $Y$ is the mass fraction of helium.

In the region (the ``ionization zone'') immediately behind the 
shock front with a characteristic length of a few ionization 
lengths, the atoms are collisionally ionized.  
Since radiative processes in this region
remove a negligible fraction of the thermal energy of the hot plasma,
this part of the shock is called the ``adiabatic'' or ``non-radiative''
zone.  Given sufficient time the radiative losses will remove the
thermal energy and cause a runaway thermal collapse of the shocked gas.
In order to maintain pressure equilibrium across the shock 
in this ``cooling region,'' an increase in density will accompany 
with a temperature decrease in the region.
Radiative cooling in the gas continues until its temperature
($T \approx 10^4 \K$) is too low for the excitation of 
UV resonance lines which are responsible for the rapid cooling.
Moreover, once the gas starts to recombine, it quickly becomes
optically thick to the ionizing radiation (mostly extreme UV
lines) produced upstream in the cooling zone.  
Roughly half of this ionizing radiation propagates downstream 
and is reprocessed into optical emission lines in this  
``photoionization zone," where the gas is maintained at a 
temperature $T_c \approx 10^4 \K$ as a result of a
balance between photoionization heating and radiative cooling.

The density of the gas in the photoionization zone is given by 
$\rho_c = 16 \rho_0 T_s / 3 T_c$.  
For a $V_s = 250 \kms$ shock, where $T_s \approx 10^6 \K$, the gas in 
the photoionization zone can be compressed by a factor $\approx 550$.  
However, a significant magnetic field entrained in the gas may 
mitigate this compression \citep{raym79}.

Shocks that have developed cooling and photoionization zones are
called ``radiative shocks.''  
A shock will be radiative if it has propagated for a characteristic 
cooling time, $t_{cool}$, which is defined as the time required for the 
shocked gas to cool from is post-shock temperature to $10^4 \K$.  
Otherwise, we call it a ``non-radiative shock.''
We calculate the cooling time $T_{cool}$ for shocks in our models 
for input shock velocities $V_s = 100 - 600 \kms$ and plot the results
in Figure~\ref{fig:cooltime}. 
We fit a power law to these results and obtain the relation
\[
t_{cool} \approx 2.2 \yrs \left( \frac{2 \times 10^4 \amucc}{\rho_0} \right)
\left( \frac{V_s}{250\kms} \right)^{3.8}\ ,
\]
where $\rho_0$ is the pre-shock density. 
Therefore, for the range of shock velocities and pre-shock 
densities expected in Spot~1, which was less than 4~years old
when our observations were made, some shocks may have 
already become radiative while others may remain non-radiative.

\subsection{Shock Emission} \label{sec-emissionlines}

After entering the shock front, atoms are collisionally ionized
until they come into equilibrium with the post-shock gas.
These ions are collisionally excited and emit line radiation,
some of which from lower ionization stages than the final 
ionization state of a species.
This ionization process in the ionization zone is not in equilibrium. 
Emission from this region lead to the detection of 
``Balmer filaments'' in other supernova remnants \citep{hest86,long90}.  
The Balmer emission comes from the shock front, where the H 
atoms are ionized.

When ions reach equilibrium with the post-shock gas, the dominant
line emission are in far UV or soft X-ray wavelengths, depending
on the post-shock temperature.  
Most of the UV and optical line emissions are produced after 
the shock starts its thermal collapse. 
While the observed high-ionization UV lines are produced in the 
cooling region, the low-ionization UV and optical lines 
are produced in the photoionization zone, which is a dense \hii\
region that is illuminated by the harder ionizing spectrum created
upstream.  

We plot the integrated surface emissivities of \nv\ \wl1240, 
\civ\ \wl1550, [\nii] \wll 6548, 6584, and \Ha\ of our shock
model described above in \S\ref{sec-shockstruc} as functions of 
downstream column density in the lower panel of Figure~\ref{fig:shock}. 
Most of these lines ($> 95\%$) are emitted during or after the 
thermal collapse of the shock, rather than in the ionization 
zone directly behind the shock front. 
Therefore, the resulting spectrum and magnitude of the 
emission from a shock depend greatly on whether the shock is
radiative or non-radiative and the cooling time of a shock 
is a good indication of the time it takes for a shock to 
``light up''.

The luminosity of a line produced by a shock in $\ergs$ is given by
\[
L = \frac{1}{2} \eta(V_s,\rho_0,t_{shock}) \rho_0 V_s^3 A_s \ ,
\]
where $\rho_0$ is the pre-shock density, $V_s$ is the shock speed,
$t_{shock}$ is the age (time since first encounter) of the shock, and
$A_s$ is the surface area that the shock covers. 
The function $\eta$ represents the fractional efficiency for a shock to
convert its thermal energy into emission for a given line. 
As shown in Figure~\ref{fig:shock}, $\eta$ can be represented as
a step function which turns on at $t_{shock} = t_{cool}$.  
We show the relation between the emission efficiency $\eta$ and
shock velocity $V_s$ for several emission lines in Figure~\ref{fig:eta}.
The thick lines in Figure~\ref{fig:eta} represent shocks that 
have completely cooled, that is, $t_{shock} \gg t_{cool}$, while 
the thin lines are 
shocks with $\tau = 3 \times (2 \times 10^4 \amucc/ \rho_0) \yrs$.  
Shocks with velocities $V_s > 250 \kms$ 
have not yet cooled by $t_{shock} = 3$~yrs and therefore
their structures are truncated when compared to the fully 
developed radiative shocks.
This shows again that radiative shocks are far more
efficient radiators than non-radiative shocks. 
Therefore, while non-radiative shocks may be present in 
the protrusion, their net contribution to the observed UV and 
optical emission from Spot~1 will be negligible.

For emission lines such as \nv\ \wl1240 (formed in the cooling region), 
and \Ha\ (formed primarily in the photoionization zone), 
the line efficiency $\eta$ has little dependence on the pre-shock
density $\rho_0$.
In contrast, forbidden lines are subject to collisional supression 
and therefore their emissivities are sensitive to $\rho_0$.  
The effect of supression is illustrated in Figure~\ref{fig:eta} by 
the [\nii] \wll6548, 6584 lines, which have a critical 
electron density of $\sim 10^5 \pcc$ \citep{oste89}. 
Effects from collisional supression increases as the shock velocity 
increases because faster radiative shocks generate more compression.
Figure~\ref{fig:eta} also shows that line emissions such 
as \nv\ \wl1240 can be produced only if the temperature of the 
shocked gas is high enough for that ion to be produced, 
as previously discussed in \citet{mich00}. 
Above that threshold, the total line emissivity increases
linearly with the shock velocity. 
Permitted emission lines formed in the photoionization zone 
have a stronger dependence on shock speed,
since they are generated by the reprocessing of the ionizing photons
produced upstream in the cooling layer. 
For example, the \Ha\ emissivity for fully developed radiative shocks 
increases approximately as $V_s^{2.4}$.

\subsection{Observed Shock Velocities and Pre-Shock Densities}

Since the gas is relatively cool ($T \approx 10^4 \K$) in the
photoionization zone, the optical lines have thermal widths of
order $10 A^{-1/2} \kms$, where $A$ is the element's atomic mass.
Macroscopic motion of the cooled layer will cause the observed line
profiles of the optical lines to be significantly broadened.
As the blast wave wraps around a protrusion, we expect to observe 
velocity components traveling both toward and away from us. 
The measured widths of the optical lines in the early \hst/STIS
spectrum suggests that the fastest radiative shock has a projected 
velocity $\sim 250 \kms$ \citep{mich00}.
Moreover, the observed line ratios of Spot~1
indicate that lower velocity ($\lae 135 \kms$) shocks must also
be present (cf. \S\ref{sec-oneshock}). 
Following the discussion in \S\ref{sec-velrange}, we are not
surprised that a range of shock velocities is required to explain the
observations.

Figure~\ref{fig:phasespace2} shows the boundaries separating 
radiative and non-radiative shocks with $t_{cool}$ of
2 and 4 years overlayed on the hydrodynamically allowed regions 
of phase space for different values for the blast-wave pressure.  
In order for shocks with velocities as low as $135 \kms$ to be 
radiative, the density of the obstacle 
$\rho_0 \gae 3 \times 10^4 \amucc$ has to be higher than 
most of the gas in the equatorial ring. 
A lower $\rho_0$ is needed if the blast-wave pressure 
is lower than our nominal estimate.  
While faster shocks propagating into less dense gas may also 
be present in Spot~1, they do not contribute significantly to the 
observed optical and UV spectra because they are non-radiative.

\section{Analysis} \label{sec-anal}

\subsection{Nebular Analysis} \label{sec-nebanal}

The \hst/STIS spectrum of Spot~1 consists of many forbidden 
emission lines of various species in the UV and optical wavelengths
(cf. Table~\ref{tb-optflux} and \ref{tb-uvflux}).
A standard nebular analysis on these data provides a good
starting point because it indicates the typical values
for the basic physical quantities that can be expected in the
line-emitting region.
We decide to focus our attention on the data taken around
1999 September ($\sim$~day 4570) when a complete set of spectrum
is obtained from $1150 - 10270$~\AA.
For each ion we consider a five-level model and include all 
relevant atomic processes, such as collisional excitation 
and de-excitation, and spontaneous radiative transitions.
The atomic data of \citet{oste89} are used for the analysis, 
except for \niv, where the results from \citet{ramsb94} are used.
We construct a grid of line ratios as functions of
electron number density $n_e$ and temperature $T_e$. 
Figure~\ref{fig:nebanal} shows the contours 
for which ratios of lines of [\oi], [\oiii], [\nii], [\sii], and 
[\siii] agree with the observed values.  

The observed ratios of various forbidden lines require a 
relatively high electron density, $n_e \approx 10^6 \pcc$,
and temperature, $T_e \approx 10^4 \K$. 
The upper limits for the line ratios of [\nI] ($5199+5201/3467 <
0.11$) and [\oii] ($3726+3729/7320+7330 < 0.3$) are consistent
with the derived high $n_e$, as well as the \niv] ($1483/1486$)
line ratio. 
Recalling that the typical number density of the ring is 
$n \sim 10^4 \pcc$ \citep{lund96}, our result indicates that the 
gas in Spot~1 must have been compressed by a factor much greater than 4.
This is consistent with our general picture (\S\ref{sec-interp})
that the UV and optical line emissions from Spot~1 originate 
from a highly compressed gas behind a radiative shock.
Our results also suggest the temperature and density stratification of       
the region where these emission lines are formed. 
This is also consistent with our general picture for radiative shocks 
where emission lines from different ionic species are created in 
different regions of the shock.

As discussed earlier (\S\ref{sec-emissionlines}), the optical 
emission lines from a radiative shock comes from a photoionization 
zone in the cooled post-shock gas.  
This photoionization picture is supported by the observed
Balmer decrement in Spot~1, \Ha\ : \Hb\ : \Hg\ : H$_{\delta} =
3.63 (\pm 0.21) : 1 : 0.48 (\pm 0.06) : 0.24 (\pm 0.04)$, which is
generally consistent with both Case A and Case B values typically 
seen in photoionized gaseous nebulae \citep{oste89}.
A complete analysis of the H and He lines will be presented in
Pun et al.\ (2002, in preparation).

\subsection{UV Emission Line Modeling} \label{sec-shockmodels}

Emission lines from the higher ionization stages 
(e.g., \niv, \civ, \oiv, \nv) are formed in the cooling region of 
a shock, 
where the gas temperature and density are stratified and the 
ionization levels can be out of equilibrium.  
However, within each ionization stage, the population of 
all energy levels are in equilibrium. 
We model these lines with a radiative shock code 
that properly account for these effects.  
We compare our model results with the observations of six UV 
emission lines detected with the G140L grating (\nv\ \wll1238, 1243, 
\SiIV\ \wl1393, 1403, \oiv] \wl1400 multiplet, \niv] \wl1486,
\civ\ \wll1548, 1543, and \heii\ \wl1640).
We choose these lines because they have negligible contamination 
from the ring emission, and because they have been observed at 
three different epochs from day~3869 to day~4596.
We decide not to model the \mgii\ \wll2796, 2803 doublet, 
the strongest emission line in the Spot~1 spectrum, because of the large 
uncertainties in the observed flux (over a factor of 2, cf.\ Table~\ref{tb-uvflux}), 
caused by huge interstellar \mgii\ absorption along the LMC line of sight
(cf. \S\ref{uv-lines}). Moreover, with the \mgii\ line flux measured
only one epoch, we are not able to monitor its time evolution as for
the other FUV lines.
Similarly, we have not attempted models to fit both the UV and
the optical emission lines because of the limited time
and wavelength coverage of the optical data (cf. Table~\ref{tb-optflux}). 

For a given set of input shock-model parameters
(shock velocity, pre-shock density, and abundances), we follow the
evolution of the downstream gas until the shocked gas has cooled 
to 5000~K.  
We store the integrated flux for various emission lines at
intermediate grid points behind the shock as the basic vectors by
which we compare with the observed data. 
We then vary the parameters in the models to obtain the best fit.
To seek the best fit, we allow the abundances to vary from our 
standard ``Ring" abundance described in \S\ref{sec-shockstruc}.

\subsubsection{Single-Shock Models} \label{sec-oneshock}

Using a grid of single-shock models with $V_s = 100 - 400 \kms$, 
we were unable to reproduce the
emission-line fluxes and widths observed from Spot~1
at any epoch.  
One of the main challenges for the single-shock model
is to explain the relatively low ratios of \nv\ \wl1240 flux to the
fluxes of lines from lower ionization stages (e.g., \niv] \wl1486 and
\civ\ \wl1550) in light of the widths ($\approx 250 \kms$) of the
observed lines.  
The abundance of \nv\ increases rapidly for shock
velocities exceeding $\approx 135 \kms$, while \niv\ and \civ\ are
present at lower shock velocities (cf. Figure~\ref{fig:eta}). 
As a result, the model line ratios 
\nv\ \wl1240/\civ\ \wl1550 $\approx 10$ 
and \nv\ \wl1240/\niv] \wl1486 $\approx 8$ for shocks faster 
than $135 \kms$ exceed the observed ratios
of $0.6 - 1.7$ and $2.5 - 5.6$, respectively.
The fact that this discrepancy exists for both line ratios
indicates that this result does not depend
on our choice of abundances.
We therefore conclude that a single-shock model cannot reproduce
the observed line emissions of Spot~1, as in 
our earlier studies \citep{mich00}.

\subsubsection{Two-Shock Models} \label{sec-twoshock}

We are not too surprised that single-shock models fail to fit
the observed emissions from Spot~1 because we expect there exists 
a range of shock velocities $V_s$ in the interaction 
(cf. \S\ref{sec-velrange}).  
The exact distribution of $V_s$ depends on the geometry and density 
distribution in the obstacle, which are both unknown.  
Rather than attempting to explore the vast parameter 
space of possible shock velocity distributions, we fit the observed 
line emissions with models consisting of only two distinct shocks, 
each having a different velocity and surface area.  
We explore parameter space with shock velocities ranging 
from 100 to $400 \kms$ and find reasonable fits to the relative 
line strengths with a combination of one ``slow'' shock with 
$V_s = 135 \kms$ and one ``fast'' shock with $V_s = 150$ -- $400 \kms$.  
As discussed below in \S\ref{sec-lineprof}, the observed line
profiles suggest that the fastest radiative shocks have velocities 
$V_s \approx 250 \kms$.
We therefore assume this value for the fast shock component.

The hydrodynamic simulations shown in Figure~\ref{fig:globalhydro} of
\S\ref{sec-globalhydro} illustrate two possible scenarios to account
for the observed increase of the line fluxes with time.  
The crucial parameter distinguishing these scenarios is the 
ratio $t_{shock} / t_{cool}$, where $t_{shock}$
is the ``age" of the interaction (i.e., the time since first impact),
and $t_{cool}$ is the characteristic radiative cooling time for the
transmitted shocks.  
For the case where both the blast-wave pressure and the obstacle density
are low, then $t_{shock} \approx t_{cool}$ and the shocks in
the obstacle would be in the process of developing their radiative
layers. 
In this scenario the observed increase of the line fluxes is
due primarily to development of new radiative layers as the shocks age. 
Alternatively, for the case where the density of the obstacle is
high, then $t_{shock} >> t_{cool}$ and the shocks quickly become radiative.
The observed increase of the line fluxes is then due to the 
increase in the surface area of the shocks as the blast wave overtakes 
more of the obstacle. 
We suspect that we are observing a combination of these two behaviors.  
Given our limited data set (6 lines in 3 epochs), we decide to 
fit only two limiting models which probably bracket the
actual situation: evolution due solely to shock aging (Model~1), and
evolution due solely to increasing shock areas (Model~2).

For Model~1 we assume that the area of each shock remains constant
over time and the only fitting parameters are the surface area and age 
of each of the two shocks.  We assume that the faster shock is
older because it represents the head-on shock that appears when the 
blast wave first encounters the obstacle, and that the slower shock is younger
because it is driven into the sides of the obstacle by an oblique
blast wave at a later time.  
For this model we need to assume a low pre-shock density 
$n_0 = 10^4 \pcc$ so  that $t_{cool} \approx t_{shock} \approx 3 \yrs$
for the fast shock.  
We also assume that the pre-shock density is constant throughout 
the obstacle for both models.
The best-fit model for Model~1 has a $\chi^2_r \approx 1.6$
and its parameters are listed in Table~\ref{tab:models}.
The model ({\it squares\/}) and observed ({\it error bars\/}) fluxes 
for all the 6 emission lines are plotted in Figure~\ref{fig:models}.
For the best-fit model, the surface area of the slower shock needs 
to be $\sim 4$ times greater than that of the faster shock to fit 
the low observed \nv\ \wl1240/\civ \wl1550 flux ratio.  
In this model, the fast shock was 2.2~years old at the time of
the first UV observation (day~3869), while the slower shock 
was only one week old.  
The age of the fast shock is only slightly smaller than the
results of \citet{lawr00a}, who suggest that Spot~1 
was at least 2.6~years old by the time of our first UV 
observation. 
This pleasant result is offset by the suspiciously
young age of the slower shock. 

For Model~2 we assume that the pre-shock density is high enough so
that all the shocks are fully radiative at all epochs.  
The fitting parameters more this model are the areas of the 
two shock surfaces at each of the three observation epochs.  
We find that the quality of the fits is rather insensitive
to the assumed pre-shock density ($n_0 = 3.3 \times10^4 \pcc$ is assumed).
This is true because the 6 UV lines we are fitting are not subject to 
collisional supression (including \niv] \wl1486, 
for which $n_{crit} > 3 \times 10^9 \pcc$), so their emission
efficiencies, $\eta$, do not depend on the pre-shock density 
(cf. \S\ref{sec-emissionlines}).
The luminosities of such lines are actually proportional to the 
product, $n_0A$, of the pre-shock density and shock area.  
We present the best-fit paramters of the model in 
Table~\ref{tab:models} and the model line fluxes in 
Figure~\ref{fig:models} ({\it circles\/}).
Since Model~2 provides a better statistical fit ($\chi_r^2 \approx 1$)
to the observed line strengths than Model~1, we conclude that the increase 
of shock areas is the dominant cause of the increase of the fluxes.

Figure~\ref{fig:area} shows the time dependence of the fitted shock
areas (actually the product $n_0A$) for Model~2.  
Similar to Model~1, we find that $\sim 4$ times more surface area 
must be covered by the slow shocks than by the fast shocks. 
While our best fit model results are statistically consistent 
with a constant ratio of slow to fast shock area 
($A_135:A_250$ in Table~\ref{tab:models}), they 
suggest that this ratio may be decreasing with time.
This behavior is not what we would expect from simple
protrusion models shown in Figure~\ref{fig:globalhydro}.  
In such models the fast shocks are created first and 
only the area of slow shock interaction increases as the blast 
wave overtakes more of the obstacle.  
One possible explanation is that there may be many
protrusions or clouds distributed in Spot~1 and therefore 
the net area of fast shocks can increase with time as more of 
them are encountered by the blast wave.
Another possible explanation arises from the fact that Model~2 
by construction does not allow for any aging of shocks.
However, the real light curve of the Spot~1 emission is probably a
combination of increased shock areas and additional emission from
older, but newly cooled, radiative shocks. 
The addition of fast shock area at later times seen in Model~2 
can therefore be attributed to shocks that were created before 
our first observation but added their emission later after 
they had cooled and become radiative.

Our finding that more area is covered by the slow shocks than fast
shocks suggests that the obstacle may have an elongated shape with 
an aspect ratio $\approx A_{slow} / A_{fast}$, where $A_{slow}$ and
$A_{fast}$, are the shock areas for the slow and fast shocks, respectively.
Assuming a simple, hemispherical geometry for the obstacle, 
we estimate that it has a typical scale length of  
$d_{spot} \approx [2 (A_{slow} + A_{fast}) / \pi]^{1/2}$.  
We find a best-fit size of the obstacle 
$d_{spot} \approx 2 \times 10^{16} (3.3 \times 10^{4} \pcc/n_0)^{1/2} \cm$ 
at day~4606, the time of our last STIS observation.  
This is consistent with our finding (cf. \S\ref{sec-obs})
that Spot~1 remains spatially unresolved in our data.

While the two-shock models successfully fit the time evolution of 
the UV emission lines, these models fail to account for the observed 
fluxes of the optical lines (e.g., \Ha, \Hb, [\nii] \wll6548, 6583, 
[\oi] \wll6300, 6364, and [\oiii] \wll4959, 5007) emitted in 
the photoionization zone (\S\ref{sec-nebanal}).
The observed optical fluxes are typically factors of $2-3$
greater than those predicted by our models.  
We exclude emission from the precursor as a possible 
contributor to the extra emission because its emission would 
be narrow and center at zero velocity. 
However, we exclude that portion of the line
profile when we fit the Spot~1 medium resolution G750M data
(cf. Fig.~\ref{fg-gaussfit}).
We discuss other possible
explanations for this discrepancy in \S\ref{sec-disc}.

Finally, we emphasize that the satisfactory fits of the UV line fluxes
were possible only if we allow abundances to vary from our
standard "Ring" abundance. The inferred
abundances are roughly consistent with the ring abundances inferred by
\citet{lund96}, but differences as large as a factor of two are
derived. 

\subsection{Line Widths and Profiles} \label{sec-lineprof}

In this section we discuss what can be learned from the widths and
profiles of the observed emission lines.  The data are sparse.  
We measure line widths of Spot~1 accurately for only a few optical 
lines (cf. \S\ref{opt-width}).
We also only have rough estimates of the widths of a few
far UV lines (cf. \S\ref{uv-flux}).

The line widths and profiles are dominated by 
the shock dynamics, and not thermal broadening.  
The line-emitting gas behind a radiative shock travels at 
approximately the shock speed and will therefore emit line 
radiation with a Doppler shift determined by the line-of-sight 
velocity of the shock.  
For emission lines originated from the photoionization zone
($T \approx 10^4$ K), e.g., the optical forbidden lines, 
the thermal broadening is $\sim 10 A^{-1/2} \kms$, which is 
much smaller than the observed widths of the lines ($\sim 250 \kms$).  
The thermal broadening for emission lines formed in the 
cooling region, such as \nv\ \wl1240, while larger than 
that for the optical forbidden lines, is still much
smaller than the dominating Doppler broadening effects 
of the macroscopic motions of the shock.
Although it is poorly measured, the width of \nv\ \wl1240 line 
(Table~\ref{tb-uvflux}) appears to be higher than that of all
the other optical and UV emission lines, with the possible exception of \Ha.  
This result can also be explained by Figure~\ref{fig:eta}, 
which shows that the \nv\ \wl1240 emission can only be produced 
in shocks faster than $\approx 135 \kms$, while all the other lines 
can be produced by even slower shocks.

The fact that the measured widths of the [\oi], [\nii], and [\sii] 
lines are smaller than \Ha\ (Table~\ref{tb-optwidth}) indicates that the 
faster shocks produce these lines with lower efficiencies 
(compared to \Ha) than the slower shocks.  
This is caused by collisional supression of the forbidden lines, which
increases with increasing compression ratio of the photoionized zone,
as shown previously in Figure~\ref{fig:eta}.

The line profile shapes depend on the geometry 
of the shock surface and its orientation relative to the observer. 
The shocks having the highest projected velocities toward and away 
from the observer produce the blue and red wings of the lines, respectively.  
The Doppler shifts of the wings of the line profile therefore 
provide a lower limit to the velocity of the fastest radiative 
shocks present while both fast transverse shocks and slow shocks
contribute to the cores of the line profiles.

We construct simulated \Ha\ line profiles based on simple geometries 
for the protrusion (e.g., Figure~\ref{fig:globalhydro}) in order to 
derive constraints on the hydrodynamics model of the interaction.
The 1999 August 30 (day~4568) G750M 0\farcs1 line profile of \Ha\ 
is chosen because it is well-observed and is not affected by 
collisional supression. 
Figure~\ref{fig:linegeom} shows our model shock surface geometries.
For each model we assume a distribution of shock velocities
(from 100 to $250 \kms$) normal to the surface of the protrusion.  
We normalize each velocity distribution by requiring that the 
relative areas of fast ($V_s > 135 \kms$) and slow ($V_s < 135 \kms$) 
shocks in each model is $\approx 4$, as determined from the 
best-fit two-shock models (\S\ref{sec-twoshock}).
We also assume that the axis of symmetry of the protrusion points 
radially inward from the circumstellar ring at the location of 
Spot~1 for all models.  
We calculate line profiles by convolving the surface 
emissivity ($\propto V_s^{2.4}$, cf. \S\ref{sec-emissionlines}) 
of the radiative shocks with the fractional areas covered by 
shocks having the appropriate line-of-sight velocities. 
We then convolve the model profile with the STIS LSF \citep{leit00}. 
At the \Ha\ wavelength, the STIS LSF for the G750M 0\farcs1
aperture can be well approximated with a Gaussian of FWHM 1.5 pixels.
The simulated profiles are finally normalized for comparison
with the observed \Ha\ profile.  

Figure~\ref{fig:lineprof} shows the resulting \Ha\ line profiles 
({\it solid\/}) from the geometry models of Figure~\ref{fig:linegeom}.
For each case the day~4568 \Ha\ profile ({\it squares\/}) is 
also plotted for comparison. 
The observed line width indicates that radiative shocks are
present with projected velocities $\approx 250 \kms$, which  
is a good indication of the velocities of the fastest radiative 
shocks present.
The line profile formed from an ellipsoidal model surface (model A) 
has multiple peaks and steep edges that are not apparent in
the observed profiles.  
The peaks are due to caustics in velocity space that occur where 
the shock surface lies tangent to the plane of the sky.  
These features are less pronounced in the line profiles
produced by a spherical model surface (model C),
where tangential surfaces comprise of a lower fraction of
the total surface areas.

For each case we also show a model in which no radiation comes from
the tip of the protrusion (models B and D, for elliptical and spherical
surfaces, respectively).  
These models are motivated by the observations from the 
{\it Chandra X-Ray Observatory\/} \citep{burr00,park01}, 
which show a bright spot in the X-ray image of \sn\ coincident with Spot~1.  
One possible interpretation is that the X-rays come from a faster 
($V_s > 600 \kms$) non-radiative shock at the tip of the protrusion 
responsible for Spot~1.
We notice that this picture is also consistent with one where
there is a density gradient across the ring. 

While the model line profiles have similar widths as the 
observed \Ha\ line, they do not provide a good fit in the
high velocity portion ($\gae 100 \kms$ and $\lae -150 \kms$) 
of the line profile.
As we shall discuss in \S\ref{sec-disc}, radiative shocks are 
subject to instabilities that will introduce ``turbulence''
in the radiative layer.  
We simulate this effect of this turbulence qualitatively by 
convolving the derived model line profiles with Gaussians of 
FWHM $100 \kms$.  
The resulting line profiles ({\it dashed lines\/} in 
Figure~\ref{fig:lineprof}) provide substantially better matches 
to the observed \Ha\ line.

\section{Discussion} \label{sec-disc}

While the idealized radiative shock models described in 
\S\ref{sec-interp} and \S\ref{sec-anal} provide a plausible
framework for interpreting the spectrum of Spot~1, they do not 
account for all the details of the observed line ratios and profiles.  
In this section, we consider what the departures of the actual spectra 
from the model spectra may tell us about how the actual physical
situation differs from our models.

First, we discuss the ``turbulent'' broadening that we 
introduce in \S\ref{sec-lineprof} to make the model line profile 
resemble the observed one at high velocity.
This turbulence is almost certainly due in part to the cooling
instabilities that exist behind radiative shocks with
$V_s \gae 150 \kms$ \citep{chev82,imam84,inne87,stri95,wald98}. 
To illustrate the effects of these instabilities,
we construct a one-dimensional VH-1 simulation of a blast
wave driving a radiative shock into gas of constant upstream
density, $n_0 = 3.3 \times 10^4 \pcc$ and the results are shown
in Figure~\ref{fig:therm1d}.
In the figure, the upstream gas flows to the left 
with a velocity $V_{inflow} = 220 \kms$ while the radiative gas,
shown as a dark band, propagates to the right with 
a mean velocity $\sim 250 \kms$.  
The dark gray parabolas represent the positions of the actual shock 
front, which moves ahead of the radiative layer and then 
collapses back to the radiative layer over time. 
The velocity of the shock front relative to the
upstream gas varies from $\sim 100$ to $300 \kms$.  
This process repeats itself on a time scale comparable to the 
cooling time of the shocked gas.  
In our simulation, the forward shock never stabilizes
while the oscillation of the velocity of the 
radiative layer decreases as it gains mass with time.

The instabilities that exist behind radiative shock become 
more dramatic in the two-dimensional simulations, 
as described in \citet{wald98}.
We construct a 2-D VH-1 shock simulation in Figure~\ref{fig:therm2d}
to illustrate such effects. 
We seed the instability with a 5\% sinusoidal density perturbation 
of the unshocked gas in a direction perpendicular to the shock front,
with the purturbation wavelength comparable to the cooling length 
of the transmitted shock.
As the shock enters the dense gas, a Richtmyer-Meshkov instability
develops rapidly at the contact discontinuity.  After several cooling
times (and forward shock oscillations), the radiative layer is
shredded into a chaotic flow of tiny blobs by a combination of
thermal and Rayleigh-Taylor instabilities.

If the blast wave strikes the obstacle at oblique incidence, the flow
of shocked gas parallel to the contact surface will cause strong
Kelvin-Helmholtz instabilities to shred the surface even further and
will introduce a component of velocity parallel to this surface in the
radiative layer.  We can see these instabilities beginning to develop
in the last frames of Figure~\ref{fig:globalhydro}.
However, we have not attempted to model their consequences in 
our simulations of the line profiles.

Since the \hst\ only allows us to resolve structures with dimensions
$\gae 10^{17}$~cm at the distance of \sn, we have no direct
knowledge of the actual shape or the density distribution of the
protrusion gas responsible for the Spot~1 emission.  
Even in the case where it has a simple shape such as that 
shown in Figure~\ref{fig:globalhydro}, the instabilities discussed 
above will introduce turbulent structures at scale lengths 
$\sim 10^{13}$~cm, which we cannot hope to resolve with current technology.

Moreover, it is unclear whether the assumption that the protrusion gas 
has a simple shape with a smooth surface is justified.
While we know that the circumstellar ring has a range of densities 
\citep{lund96}, we do not know the scale lengths upon which these 
inhomogeneities may be distributed. 
If the blast wave encounters a ``lumpy'' medium, a complex flow 
will develop where the radiating blobs are hammered both 
by the reflected shocks and by the blast wave itself. 
Clearly, the current observations allow for many plausible scenarios 
for the hydrodynamics of the radiating gas for which quantitative 
models are beyond the scope of this paper.  
We are not surprised that the simple hydrodynamic models we have
considered in \S\ref{sec-lineprof} fail to reproduce the observed 
line profiles.

Magnetic fields may also affect the shock dynamics.  For the range of
densities and shock velocities expected in Spot 1, we estimate that
a magnetic field with strength $B_0 \gae 400 \ \mu$G entrained in the 
upstream gas would be sufficient to suppress compression in radiative 
shocks and stabilize the cooling instabilities \citep{inne92}. 
This will result in a photoionization zone of lower density.

In our idealized shock models of \S\ref{sec-interp} and \S\ref{sec-anal}, 
we have only considered radiative shocks in which the pressure 
driving the shocks is constant.  
However, as \citet{bork97b} and \citet{luo94} have shown, the pressure
behind an actual blast wave encountering an obstacle have a complex
time dependency. 
As the reflected blast wave travels away from the
surface of the obstacle the flow diverges and the driving pressure
decreases.  Later, the pressure driving the radiative shock into the
obstacle may suddenly rise again owing to ``echo" shocks produced when
the reflected shock strikes the contact discontinuity and the reverse
shock.

The most troubling failure of our models is their inability to account
for the observed ratio of fluxes of optical to UV lines.  As
we have described, the radiative shock models consistently predict
optical fluxes that are factors $\approx 3$ lower than expected from
models that are normalized to give the observed UV line
fluxes.

A simple way to account for the discrepancy between the observed
and model optical/UV line ratios is to assume that the extinction of
Spot~1 is greater than that previously determined for SN 1987A.  
If we applied a reddening of $E(B-V) \approx 0.23$, the corrected
UV flux would increase by a factor $\approx 3$, sufficient 
to remove the discrepancy.  
Considering that Spot~1 is located on the far side of the ring
from the observer, a column density $n \ell \approx 10^{21} \col$ of gas
with a 30 Doradus dust/gas ratio in the foreground of Spot~1 (perhaps
the ring material itself) could account for such additional reddening.
It is also possible that the shock itself may destroy graphite 
dust grains and elevate the ratio of C/N abundances in Spot~1 
(compared to the unshocked ring) to account in
part for the relatively low observed ratio of \nv\ \wl1240/\civ\ \wl1550.
Shocks faster than $\approx 200 \kms$ can destroy graphite
grains in a timescale $\lae 2 \yrs$ \citep{tiel94}.

The revised value of $E(B-V)$ is 0.04 (and $2\sigma$) greater than 
the measurement of $0.19 \pm 0.02$ by \citet{scud96} on the nearby Star~2, 
and 0.07 greater than the value we adopted for Table~\ref{tb-uvflux}. 
However, modeling of the UV and optical narrow emission lines 
for the inner circumstellar ring by \citet{lund96} does not
support such large $E(B-V)$ value.
Therefore we do not regard it to be highly probable.

A second way to account for the optical/UV line ratio is 
the presence of an extra source of ionization in the photoionization 
zone that is roughly three times stronger than the radiation from 
the shocked gas itself.
In the one-dimensional radiative shock models described in \S\ref{sec-interp}, 
the only source of ionizing radiation for the photoionized zone 
is the shocked gas immediately upstream from that layer.  
One possible candidate for the extra ionization is the faster shocks 
which have not had time to develop their own radiative layers.  
However, we are not confident that such models will work.  
The problem is that the presumed extra ionizing radiation must
also illuminate the cooling zone where the UV emission lines
are formed, and this illumination will elevate the abundance of \nv\
in the cooling layer.  As a result, shocks with velocity 
$V_s \lae 135 \kms$ will produce extra \nv\ \wl1240 emission, 
causing the predicted \nv\ \wl1240 flux to again exceed 
the observed value.
A more promising candidate for the extra source of
ionization is the nonthermal radiation from particles accelerated 
in the shock itself. 
Detailed modeling is required to demonstrate its feasibility.

A third possible explanation for the failure of the models 
to account for the ratios of optical to UV emission lines is the
inadequacy of the stationary radiative shock models themselves.  
Our findings are similar to that from the modeling of the spectra 
of Herbig-Haro objects with simple plane-shock and bow-shock models,
where high-ionization lines, such as \civ\ and \nv, are underproduced
\citep{bohm97}.
As we have shown in Figure~\ref{fig:therm2d}, radiative shocks are
inherently unstable and the ratio of optical to UV emission
lines from such unstable shocks may be different from that in a
stationary flow, even in a time-averaged or ensemble-averaged sense.
For example, the temperature gradients in the unstable flow are
more pervasive than in a stationary flow, in which case thermal
conduction may provide an additional energy source.

Future observations with the \hst\ may be very helpful in resolving 
some of the issues that we have raised here.
Valuable clues to resolving the optical/UV line ratio puzzle may
come from observations of the same ratio in the other hot spots.
For example, if the ratio is affected
by extinction by dust in the immediate foreground of Spot~1, it might
have different values in the other spots, some of which lie on the
near side of the circumstellar ring.

We can also benefit from observations of the UV emission lines at a 
higher spectral resolution.  
First, we can better estimate the amount of flux reduction caused 
by the LMC interstellar absorption (cf. \S\ref{uv-flux}) 
by obtaining the line profile shapes of emissions not affected by 
this absorption, such as \nv\ \wl1240, and \niv\ \wll1483, 1486.
This will significantly reduce the error estimates of the measured
fluxes affected by this absorption, especially for \mgii\ \wl2800, 
where the flux correction due to LMC absorption applied is both the 
highest and the most uncertain. 
Second, to reconcile the observed low flux ratio 
of \nv\ \wl1240/\civ\ \wl1550, 
we have suggested a two-shock model in which
the higher-velocity radiative shocks make up a relatively small
fraction of the shock area compared to the lower-velocity shocks.
This model implies that the ratio of \nv\ \wl1240/\civ\ \wl1550 line
emissivity should increase in the wings of the lines, where only
the fastest shocks contribute.  
On the other hand, we have also suggested that destruction of 
graphite grains in the faster shocks may elevate the abundance of 
carbon in the gas phase and lead to a decrease in the ratio of 
\nv\ \wl1240/\civ\ \wl1550 in the wings of the lines.  
However, as shown in Figure~\ref{fg-absorb}, the observed profile of 
\civ\ \wl1550 is severely modified by interstellar absorption,
introducing large uncertainties in the N/C abundance ratio. 
We may get a clear look at the red wing of \civ\ \wl1550
by observing the spots on the far side of the ring (e.g.,
Spots 3, 4, and 5), where the shock emission would be mostly
redshifted to wavelengths where the interstellar absorption is not
so severe.  When Spot~1 becomes bright enough to permit observations of
the line profile of \niv] \wl1486, the flux ratio of 
\niv] \wl1486/\nv\ \wl1240 as a function of position within the line 
profile will give us a diagnostic of shock excitation versus velocity 
that is independent of relative abundances and not
subject to interstellar absorption.

\section{Summary} \label{sec-conclusion}

In conclusion we summarize the main results of this paper:

\begin{itemize}

\item{We describe the \hst/STIS UV and optical spectrum of 
Spot~1 up to day 4606 (1999 October 7).  
We measure and tabulate fluxes, line widths, and
line centroids for all lines detected in the spectrum.}

\item{We determine correction factors to account for 
the substantial interstellar line absorption of several lines (\cii\
\wl1335 multiplet, \SiIV\ \wll1394, 1403, \civ\ \wll1548, 1551, and
\mgii\ \wll2796, 2803). However, these factors are poorly 
constrained because of uncertainties in the line widths 
and centroids.}

\item{The observed emission is caused by radiative shocks which
develop when the supernova blast wave strikes dense gas protruding
inward from the equatorial ring.  A nebular analysis of several line
ratios confirms that many of the emission lines are formed in a region
of very high density gas ($n_e \sim 10^6 \pcc$), which has been
compressed by factors $\gae 100$ due to radiative cooling downstream
from the shock.}

\item{The observed line widths indicate that radiative shocks with
velocities as high as $250 \kms$ are present, while the line ratios
indicate that much of the radiation must come from slower shocks ($V_s
\lae 135$).  The inferred range of shock velocities is a natural
consequence of a model in which a blast wave overtakes a dense
obstacle.}

\item{Hydrodynamic arguments show that such slow shock velocities will
be present only if the density in the unshocked obstacle is 
sufficiently high ($n_0 \approx 3 \times 10^4 \pcc$).  
This density lies at the high end of the range of densities 
observed for the ring by \citet{lund96}.  The fact that shocks 
with velocity $V_s = 250 \kms$ must have become radiative after 3 years 
sets a lower density limit, $n_0 \gae 10^4 \pcc$.}

\item{Faster shocks ($V_s \approx 250$ -- $1000 \kms$) may also be
present in Spot~1, as would be the case for shock interaction
with the lower-density gas ($n_0 < 10^4 \pcc$) present in the ring
\citep{lund96}. These shocks would be invisible in the
optical and UV emission spectra though because they would not yet have 
developed radiative layers.  
However, they may contribute to the enhancements of
X-ray emission seen near the hot spots by the 
{\it Chandra X-ray Observatory\/} \citep{burr00,park01}.}

\item{We interpret a subset of the observed emission lines (6 UV
lines observed at 3 epochs) with a model consisting of two
one-dimensional stationary radiative shocks with different velocities
($V_s = 135$ and $250 \kms$).  In our favored model, the lines brighten
because the area of the obstacle overtaken by the blast increases with
time.  The area covered by the slower shocks must be $\approx 4$ times
that covered by the faster shocks.  To account for the observed line
strengths, the obstacle must have a characteristic scale length of
$2 \times 10^{16} \cm$.}

\item{Our models underestimate the ratio of observed
optical/UV line fluxes by a factor of $\sim 3$. 
This may indicate the presence of an additional 
source of ionizing flux in the photoionization zone,
or that the assumption of steady state shocks is too limiting.}

\item{The observed line profiles are in qualitative agreement with
simulated profiles created from simple geometries for the obstacle.
The nearly Gaussian observed profiles indicate the presence of chaotic
flows in the photoionized region, caused by violent thermal and shear
instabilities in the shocked gas.}

\item{Continued UV and optical observations of Spot~1 are necessary 
to further our understanding of the physics of the radiative 
shocks present there.
Resolving the profile of the \nv\ \wl1240 and other UV lines
will provide critical tests of the radiative shock model.}

\item{Observations and modeling of the emission-line spectra of the
other spots now appearing in the ring will provide vital clues to
resolving the uncertainties of the present model for Spot~1.
Observations with \hst/STIS are required to spatially resolve their
individual spectra.}

\end{itemize}

\acknowledgments We thank Dan Welty for sharing with us
the high resolution \iue\ observations of \sn, 
Kailash Sahu for sending us the most recent STIS LSF, 
John Raymond for providing us with his radiative shock code, 
and John Blondin for use of his VH-1 code and his help with 
hydrodynamic simulations for this paper.
Support for this work was provided by NASA through grant GO-08243
from the Space Telescope Science Institute, which is operated by
the Association of Universities for Research in Astronomy Inc., under
NASA contract NAS5-26255.
Additional support was provided by NASA through grants NAG5-3313
and NTG5-80 to the University of Colorado.
C.S.J.P. acknowledges funding by the STIS IDT through the National
Optical Astronomy Observatories.
A.V.F. is grateful for a Guggenheim Foundation Fellowship.

\clearpage
\begin{deluxetable}{lccclcr}
\tabletypesize{\footnotesize}
\tablewidth{0pc}
\tablecaption{
{\it HST\/}/STIS observations of Spot~1. \label{tb-obs}}
\tablehead{
\colhead{Grating} & \colhead{} & 
\colhead{Wavelength} & \colhead{Resolution} &
\colhead{} & \colhead{Days after} &
\colhead{Exposure} \\
\colhead{setting} & \colhead{Slit} &
\colhead{Range (\AA)} & \colhead{(\AA)} &
\colhead{Date} & \colhead{outburst} &
\colhead{time (sec)}}
\startdata
G140L (1425) & $52\arcsec\times0\farcs5$ & $1140 - 1700$ & 1.0 & 1997 September 27 & 3869.3 & 11222 \\
G140L (1425) & $52\arcsec\times0\farcs5$ & $1140 - 1700$ & 1.0 & 1999 February  27 & 4387.6 & 14350 \\
G140L (1425) & $52\arcsec\times0\farcs2$ & $1140 - 1700$ & 1.0 & 1999 October 7    & 4606.1 & 10478 \\
             &               &               &     &                   &        &       \\
G230L (2376) & $52\arcsec\times0\farcs2$ & $1570 - 3180$ & 3.0 & 1999 September 17 & 4586.5 & 10125 \\
             &               &               &     &                   &        &       \\
G430L (4300) & $52\arcsec\times0\farcs2$ & $2900 - 5700$ & 4.0 & 1999 September 3  & 4571.9 &  7583 \\
             &               &               &     &                   &        &       \\
G750L (7751) & $52\arcsec\times0\farcs5$ & $5240 - 10270$& 8.0 & 1999 February 21  & 4380.9 & 10500 \\
G750L (7751) & $52\arcsec\times0\farcs2$ & $5240 - 10270$& 8.0 & 1999 September 18 & 4387.4 &  7583 \\
             &               &               &     &                   &        &       \\
G750M (6581) & $52\arcsec\times0\farcs2$ & $6295 - 6867$ & 1.0 & 1998 March 7      & 4030.0 &  8056 \\
G750M (6581) & $52\arcsec\times0\farcs1$ & $6295 - 6867$ & 1.0 & 1999 August 30 & 4368.0 & $3\times7804$\tablenotemark{a} \\
\enddata
\tablenotetext{a}{Not all observations centered on Spot~1 (refer to text).}
\end{deluxetable}

\clearpage
\begin{deluxetable}{llrrrrc}
\tabletypesize{\scriptsize}
\tablewidth{0pc}
\tablecaption{{\it HST\/}/STIS dereddened optical emission-line fluxes from Spot 1 \label{tb-optflux}}
\tablehead{
\colhead{} & \colhead{Central} & 
\colhead{1998 Mar 7} & 
\colhead{1999 Feb 21} &
\colhead{1999 Sep 1\tablenotemark{a}} &
\colhead{1999 Sep 18} &
\colhead{} \\
\colhead{Emission} & 
\colhead{Wavelength} & 
\colhead{(day 4030.0)} & 
\colhead{(day 4380.9)} & \colhead{(day 4570.0)} & \colhead{(day 4587.4)} &
\colhead{Extinction} \\
\colhead{} & \colhead{(\AA)} & 
\colhead{flux\tablenotemark{b}} &
\colhead{flux\tablenotemark{b}} &
\colhead{flux\tablenotemark{b}} &
\colhead{flux\tablenotemark{b}} & 
\colhead{correction\tablenotemark{c}}}
\startdata
\lbrack\ion{Ne}{5}]& 3425.8 & \nodata       & \nodata       & $2.3 \pm 1.6$ & \nodata & 2.09 \\
\lbrack\ion{N}{1}] & 3466.5 & \nodata       & \nodata       & $7.2 \pm 1.9$ & \nodata & 2.08 \\
\lbrack\ion{O}{2}]& 3726.0 + 3728.8 & \nodata & \nodata     & $<$ 6.0       & \nodata & 2.02 \\
H$\iota$           & 3770.6 & \nodata       & \nodata       & $1.2 \pm 1.0$ & \nodata & 2.01 \\
H$\theta$          & 3797.9 & \nodata       & \nodata       & $1.6 \pm 1.2$ & \nodata & 2.00 \\
H$\eta$            & 3835.4 & \nodata       & \nodata       & $4.4 \pm 1.3$ & \nodata & 2.00 \\
\lbrack\ion{Ne}{3}]& 3868.7 & \nodata       & \nodata       & $13.3\pm 2.7$ & \nodata & 1.99 \\
\ion{He}{1} + H$\zeta$& 3888.7 + 3889.1& \nodata& \nodata   & $18.0\pm 2.8$  & \nodata & 1.98 \\
\ion{Ca}{2} + \lbrack\ion{Fe}{2}] (8F)& 3933.7 + 3931.4 + 3932.7& \nodata& \nodata& $4.6 \pm 1.0$ & \nodata & 1.97 \\
\ion{He}{1} + \ion{Ca}{2} + \lbrack\ion{Ne}{3}]& 3964.7 + 3968.5 + 3968.7 & \nodata& \nodata& $12.5\pm2.6$\tablenotemark{d} & \nodata & 1.96 \\
H$\epsilon$        & 3970.1 & \nodata       & \nodata       & $12.5\pm2.6$\tablenotemark{d} & \nodata & 1.96 \\
\ion{He}{1}        & 4026.2 & \nodata       & \nodata       & $2.3 \pm 1.0$ & \nodata & 1.95 \\
\lbrack\ion{S}{2}] & 4068.6 & \nodata       & \nodata       & $21.8\pm 2.4$ & \nodata & 1.93 \\
\lbrack\ion{S}{2}] & 4076.4 & \nodata       & \nodata       & $10.5\pm 1.3$ & \nodata & 1.93 \\
H$\delta$          & 4101.7 & \nodata       & \nodata       & $12.9\pm 1.8$ & \nodata & 1.92 \\
\lbrack\ion{Fe}{2}]& 4244.0 & \nodata       & \nodata       & $2.6 \pm 0.8$ & \nodata & 1.88 \\
H$\gamma$          & 4340.5 & \nodata       & \nodata       & $24.9\pm 2.5$ & \nodata & 1.85 \\
\lbrack\ion{Fe}{2}] (21F)& 4358.4 & \nodata & \nodata       & $4.9 \pm2.5$\tablenotemark{d} & \nodata & 1.84 \\
\lbrack\ion{O}{3}] & 4363.2 & \nodata       & \nodata       & $4.9 \pm2.5$\tablenotemark{d} & \nodata & 1.84 \\
\ion{He}{1}        & 4387.9 & \nodata       & \nodata       & $1.2 \pm 0.5$ & \nodata & 1.84 \\
\lbrack\ion{Fe}{2}] (6F)& 4416.3& \nodata   & \nodata       & $3.2 \pm 0.7$ & \nodata & 1.82 \\
\lbrack\ion{Fe}{2}] (7F + 6F)& 4452.1 + 4458.0 & \nodata  & \nodata       & $2.0 \pm 0.7$ & \nodata & 1.81 \\
\ion{He}{1} + \lbrack\ion{Fe}{2}] (6F)& 4471.5 + 4470.3& \nodata& \nodata & $4.6 \pm 0.7$ & \nodata & 1.81 \\
\ion{Mg}{1}]       & 4571.1 & \nodata       & \nodata       & $4.1 \pm 1.0$ & \nodata & 1.78 \\
\lbrack\ion{Fe}{2}] (4F + 5F)& 4664.5,4665.7 + 4665.0 & \nodata& \nodata  & $2.0 \pm 0.8$ & \nodata & 1.75 \\
\ion{He}{2} + \lbrack\ion{Fe}{2}] (5F)& 4685.7 + 4687.6 & \nodata& \nodata& $4.0 \pm 0.7$ & \nodata & 1.75 \\
\lbrack\ion{Fe}{2}] (20F)& 4814.6 & \nodata & \nodata       & $1.7 \pm 0.5$ & \nodata & 1.71 \\
H$\beta$           & 4861.3 & \nodata       & \nodata       & $49.6\pm 2.7$ & \nodata & 1.70 \\
\lbrack\ion{Fe}{2}] (3F + 4F)& 4889.7 & \nodata & \nodata       & $1.7 \pm 0.7$ & \nodata & 1.70 \\
\lbrack\ion{Fe}{2}] (20F)& 4905.4 & \nodata       & \nodata       & $1.4 \pm 0.5$ & \nodata & 1.69 \\
\ion{He}{1}        & 4921.9 & \nodata       & \nodata       & $0.6 \pm 0.3$ & \nodata  & 1.69 \\
\lbrack\ion{O}{3}] & 4958.9 & \nodata       & \nodata       & $7.7 \pm 1.4$ & \nodata & 1.68 \\
\lbrack\ion{O}{3}] & 5006.8 & \nodata       & \nodata       & $24.4\pm 1.1$ & \nodata & 1.67 \\
\ion{He}{1}        & 5015.7 & \nodata       & \nodata       & $4.5 \pm 0.7$ & \nodata & 1.67 \\
\lbrack\ion{Fe}{2}] (18F + 19F)& 5158.0 + 5158.8 & \nodata& \nodata       & $4.6 \pm 0.6$ & \nodata       & 1.64 \\
\lbrack\ion{N}{1}] & 5197.9 + 5200.3 & \nodata & \nodata    & $<$ 0.8       & \nodata & 1.63 \\
\lbrack\ion{Fe}{2}] (19F + 18F)& 5261.6 + 5268.9,5273.4 & \nodata& \nodata& $4.8 \pm 0.7$ & \nodata       & 1.62 \\
\lbrack\ion{Fe}{2}] (19F)& 5333.7 & \nodata & \nodata       & $1.0 \pm 0.3$ & \nodata       & 1.60 \\
\lbrack\ion{Fe}{2}] (19F)& 5376.5 & \nodata & $0.6 \pm 0.4$ & $1.4 \pm 0.3$ & $0.6 \pm 1.0$ & 1.60 \\
\ion{He}{2} + [\ion{Fe}{2}] (17F)& 5411.5 + 5412.6 &\nodata& $1.0 \pm 0.3$ & $1.5 \pm 0.4$ & $1.4 \pm 0.6$ & 1.59 \\
\lbrack\ion{Fe}{2}] (34F)& 5527.3 & \nodata & \nodata       & $1.1 \pm 0.3$ & $1.4 \pm 0.5$ & 1.57 \\
\lbrack\ion{O}{1}] & 5577.4 & \nodata       & $1.0 \pm 0.5$ & $1.9 \pm 0.4$ & $2.3 \pm 0.6$ & 1.57 \\
\lbrack\ion{N}{2}] & 5754.6 & \nodata       & $16.9\pm 2.0$ & \nodata       & $20.8\pm 1.4$ & 1.54 \\
\ion{He}{1}        & 5875.7 & \nodata       & $9.3 \pm 2.1$ & \nodata       & $14.4\pm 1.1$ & 1.53 \\
\lbrack\ion{O}{1}] & 6300.3 & $13.2\pm 1.0$ & $18.7\pm 1.5$ & $27.1\pm 1.2$ & $25.9\pm 1.9$ & 1.48 \\
\lbrack\ion{S}{3}] & 6312.1 & \nodata       & \nodata       & \nodata       & $0.6 \pm 0.3$ & 1.48 \\
\lbrack\ion{O}{1}] & 6363.8 & $4.9 \pm 0.6$ & $7.4 \pm 0.5$ & $8.1 \pm 0.9$ & $8.0 \pm 1.3$ & 1.47 \\
\lbrack\ion{N}{2}] & 6548.1 & $8.2 \pm 1.1$ & \nodata       & $9.5 \pm 1.0$ & \nodata       & 1.45 \\
H$\alpha$          & 6562.8 & $67.9\pm 1.8$ & \nodata       & $179.8\pm3.9$ & \nodata       & 1.45 \\
\lbrack\ion{N}{2}] & 6583.5 & $19.0\pm 1.3$ & \nodata       & $31.1\pm 1.7$ & \nodata       & 1.45 \\
\ion{He}{1}        & 6678.2 & $1.2 \pm 0.3$ & $1.6 \pm 0.2$ & $2.2 \pm 0.5$ & $3.5 \pm 0.6$ & 1.44 \\
\lbrack\ion{S}{2}] & 6716.5 & $1.3 \pm 1.3$ & \nodata       & $1.3 \pm 0.8$ & \nodata       & 1.44 \\
\lbrack\ion{S}{2}] & 6730.8 & $1.3 \pm 0.5$ & \nodata       & $2.2 \pm 0.7$ & \nodata       & 1.44 \\
\ion{He}{1}        & 7065.2 & \nodata       & $4.3 \pm 0.9$ & \nodata       & $6.3 \pm 0.7$ & 1.40 \\
\lbrack\ion{Ar}{3}]& 7135.9 & \nodata       & $2.6 \pm 0.4$ & \nodata       & $2.5 \pm 0.7$ & 1.40 \\
\lbrack\ion{Fe}{2}] (14F)& 7155.2 & \nodata & $3.0 \pm 0.6$ & \nodata       & $4.3 \pm 0.8$ & 1.39 \\
\lbrack\ion{Fe}{2}] (14F)& 7172.0 & \nodata & $3.7 \pm 0.7$ & \nodata       & $1.2 \pm 0.5$ & 1.39 \\
\lbrack\ion{Ca}{2}]& 7291.5 & \nodata       & $4.1 \pm 0.7$ & \nodata       & $6.2 \pm 0.9$ & 1.38 \\
\lbrack\ion{Ca}{2}] + [\ion{O}{2}]& 7323.9 + 7320,7330& \nodata& $20.9\pm 4.0$ & \nodata  & $25.2\pm 1.3$ & 1.38 \\
\lbrack\ion{Ni}{2}] (2F)& 7377.9 & \nodata  & $0.4 \pm 0.5$ & \nodata       & $2.1\pm 0.9$\tablenotemark{e}& 1.37 \\
\lbrack\ion{Fe}{2}] (14F)& 7388.2 & \nodata & $0.3 \pm 0.4$ & \nodata       & $2.1\pm 0.9$\tablenotemark{e}& 1.37 \\
\lbrack\ion{Ni}{2}] (2F)& 7411.6 & \nodata  & $0.4 \pm 0.4$ & \nodata       & {\bf WEAK}    & 1.37 \\
\lbrack\ion{Fe}{2}] (14F)& 7452.6 & \nodata & $0.4 \pm 0.2$ & \nodata       & $1.1 \pm 0.5$ & 1.36 \\
\lbrack\ion{Fe}{2}] (13F)& 8617.0 & \nodata & $1.1 \pm 0.4$ & \nodata       & $4.4 \pm 0.9$ & 1.27 \\
\lbrack\ion{S}{3}] & 9068.6 & \nodata       & $0.8 \pm 0.5$ & \nodata       & $1.4 \pm 0.9$ & 1.24 \\
\lbrack\ion{S}{3}] & 9530.6 & \nodata       & $4.8 \pm 1.7$ & \nodata       & $11.2\pm 1.6$ & 1.22 \\
\enddata
\tablenotetext{a}{Combining the G750M data taken on August 30, 1999 (day 4568.0)
                  and the G430L data taken on September 3, 1999 (day 4571.9).}
\tablenotetext{b}{Flux in units of $10^{-16} \ \ergcms$. } 
\tablenotetext{c}{$E(B-V)$ = 0.16, and the extinction correction law
of \citet{card89}, with R$_{V}$ = 3.1.}
\tablenotetext{d}{Emission blended in the G430L data, separate component determined from
                  Jan 2000 data (Pun et al. 2002, in preparation).}
\tablenotetext{e}{Emissions blended in the G750L data.}
\end{deluxetable}

\clearpage
\begin{deluxetable}{lcccc}
\tabletypesize{\small}
\tablewidth{0pc}
\tablecaption{
Peak emission velocity and FWHM width of Spot 1 emissions \label{tb-optwidth}}
\tablehead{
\colhead{Emission} & \multicolumn{2}{c}{1998 Mar 7 (day 4030.0)} & 
                     \multicolumn{2}{c}{1999 Aug 30 (day 4568.0)} \\
\colhead{Line}     & \colhead{Width\tablenotemark{a}} 
                   & \colhead{Peak velocity\tablenotemark{a}} 
                   & \colhead{Width\tablenotemark{a}} 
                   & \colhead{Peak velocity\tablenotemark{a}}}
\startdata
\lbrack\ion{O}{1}] \wl6300 & 154 $\pm$ \phn7& -13 $\pm$   18 & 172 $\pm$ \phn5 & -34 $\pm$ 5 \\
\lbrack\ion{O}{1}] \wl6363 & 179 $\pm$   14 & -27 $\pm$   20 & 182 $\pm$    11 & \nodata     \\
\lbrack\ion{N}{2}] \wl6548 & 165 $\pm$   11 & -37 $\pm$   24 & 230 $\pm$    16 & -11 $\pm$ 4 \\
H$\alpha$ \wl6563          & 253 $\pm$ \phn4& -29 $\pm$ \phn3& 225 $\pm$ \phn3 & -35 $\pm$ 2 \\
\lbrack\ion{N}{2}] \wl6584 & 204 $\pm$ \phn7& -17 $\pm$   10 & 215 $\pm$ \phn7 & -19 $\pm$ 2 \\
\ion{He}{1} \wl6678        & 177 $\pm$   38 & \nodata        & 183 $\pm$  27   & \nodata \\
\lbrack\ion{S}{2}] \wl6717 & 113 $\pm$   40 & \nodata        & \phn84 $\pm$ 24 & \nodata \\
\lbrack\ion{S}{2}] \wl6731 & 149 $\pm$   48 & \nodata        & 134 $\pm$  20   & \nodata \\
\enddata
\tablenotetext{a}{Velocity in units of $\kms$.}
\end{deluxetable}

\clearpage
\begin{deluxetable}{llcrcrcrcc}
\tabletypesize{\scriptsize}
\tablewidth{0pc}
\tablecaption{
{\it HST\/}/STIS dereddened UV emission-line fluxes and line widths of Spot~1
\label{tb-uvflux}}
\tablehead{
\colhead{} & \colhead{Central} & \colhead{Flux} &
\multicolumn{2}{c}{1997 Sep 27 (day 3869.3)} & 
\multicolumn{2}{c}{1999 Feb 27 (day 4387.6)} & 
\multicolumn{2}{c}{1999 Sep 27 (day 4596.2)} & 
\colhead{Extinction} \\
\colhead{Emission} & 
\colhead{Wavelength (\AA)\tablenotemark{a}} & \colhead{correction\tablenotemark{b}} &
\colhead{Flux\tablenotemark{c}} & \colhead{Width\tablenotemark{d}} & 
\colhead{Flux\tablenotemark{c}} & \colhead{Width\tablenotemark{d}} & 
\colhead{Flux\tablenotemark{c}} & \colhead{Width\tablenotemark{d}} & 
\colhead{correction\tablenotemark{e}}}
\startdata
\ion{N}{5}  &1238.8 + 1242.8& \nodata          & $15.0 \pm 5.3$&$270^{+220}_{-\ldots}$&$51.5 \pm 6.4$&$380^{+150}_{-180}$   & $79.6\pm 5.6$ & $330^{+110}_{-140}$  & 5.93 \\
\ion{C}{2}  &1335 multiplet & 2.6$^{+1.5}_{-0.8}$& \nodata       & \nodata      & \nodata       & \nodata              &$\phn7.3\phn ^{+\phn5.4}_{-\phn4.2}$& \nodata& 4.96 \\
\ion{Si}{4} &1393.8 + 1402.8& 2.0$^{+0.8}_{-0.5}$& $\phn5.8\phn^{+\phn3.2}_{-\phn2.6}$& \nodata &$14.9\phd^{+\phn6.9}_{-\phn5.1}$&$130^{+150}_{-\ldots}$&
            $21.0\phd^{+10.9}_{-\phn8.8}$& $160^{+180}_{-\ldots}$ & 4.57 \\
\ion{O}{4}] & 1400 multiplet& \nodata          & $\phn6.1\pm1.7$& \nodata      & $11.0 \pm 2.4$& \nodata              & $11.1 \pm 2.9$& \nodata                & 4.53 \\
\ion{N}{4}] & 1483.3        & \nodata          & $\phn5.9\pm1.5$& \nodata      &$\phn6.5\pm1.3$&$110^{+130}_{-\ldots}$&$\phn9.2\pm1.4$& $200^{+100}_{-\ldots}$ & 4.16 \\
\ion{N}{4}] & 1486.5        & \nodata          & $\phn6.1\pm1.5$& \nodata      & $11.5 \pm 1.9$&$110^{+130}_{-\ldots}$& $14.2 \pm 1.8$& $200^{+100}_{-\ldots}$ & 4.15 \\
\ion{C}{4}  &1548.2 + 1550.8& 2.1$^{+1.0}_{-0.5}$& $24.1\phd^{+13.3}_{-\phn9.0}$& \nodata &$37.3\phd^{+18.9}_{-11.5}$&$150^{+100}_{-\ldots}$&
            $45.7\phd^{+23.4}_{-14.4}$& $180^{+150}_{-\ldots}$ & 3.95 \\
\lbrack\ion{Ne}{4}]& 1601.5 + 1601.7& \nodata& \nodata& \nodata      & \nodata       & \nodata      &$\phn3.9\pm2.4$& \nodata       & 3.81 \\
\ion{He}{2} & 1640.4        & \nodata          & $13.1\pm 3.4$ & \nodata      & $36.7 \pm 4.8$&$100^{+80}_{\ldots}$  & $64.3\pm 8.9$ & $130^{+80}_{-\ldots}$ & 3.73 \\
\ion{O}{3}  &1660.8 + 1666.2& \nodata          & \nodata       & \nodata      & \nodata       & \nodata      & $13.7 \pm 2.8$& \nodata       & 3.69 \\
\ion{C}{3}] & 1908.7        & \nodata          & \nodata       & \nodata      & \nodata       & \nodata      & $19.7 \pm 5.2$& \nodata       & 3.56 \\
\ion{N}{2}] &2139.0 + 2142.8& \nodata          & \nodata       & \nodata      & \nodata       & \nodata      & $18.5 \pm 3.2$& \nodata       & 4.02 \\
\ion{C}{2}  &2325 multiplet & \nodata          & \nodata       & \nodata      & \nodata       & \nodata      & $33.4 \pm 2.2$& \nodata       & 3.43 \\
\ion{Si}{2} & 2334.6        & \nodata          & \nodata       & \nodata      & \nodata       & \nodata      &$\phn6.9\pm1.5$& \nodata       & 3.39 \\
\lbrack\ion{O}{2}]& 2470.3  & \nodata          & \nodata       & \nodata      & \nodata       & \nodata      & $10.7 \pm 1.5$& \nodata       & 2.93 \\
\nodata     & 2736.7\tablenotemark{f} & \nodata& \nodata       & \nodata      & \nodata       & \nodata      &$\phn2.3\pm0.8$& \nodata       & 2.51 \\
\nodata     & 2746.3\tablenotemark{f} & \nodata& \nodata       & \nodata      & \nodata       & \nodata      &$\phn3.0\pm0.8$& \nodata       & 2.50 \\
\ion{Mg}{2} &2795.5 + 2802.7& 6.6$^{+14.2}_{-3.9}$& \nodata      & \nodata      & \nodata    & \nodata      &$230\phd^{+\phd493}_{-\phd135}$ & \nodata  & 2.45 \\
\enddata

\tablenotetext{a}{Wavelength in $\lambda_{\rm vac}$ for $\lambda <$~2000~\AA,
                  $\lambda_{\rm air}$ for $\lambda >$~2000~\AA.}
\tablenotetext{b}{Corrected for LMC line absorption assuming Spot~1 FWHM 
                  velocity $V_{\rm FWHM} = 150 ^{+ 150}_{-\ 50} \ \kms$ and peak velocity
                  $V_{0} = -30 \pm 15 \kms$.}
\tablenotetext{c}{Flux in units of $10^{-16} \ergcms$.} 
\tablenotetext{d}{Emission-line widths in units of $\kms$.} 
\tablenotetext{e}{$E(B-V)_{\rm LMC} = 0.06$ and $E(B-V)_{\rm Galactic} = 0.10$ are
                  assumed, along with $R_V = 3.1$. 
                  For the LMC component, the 30 Doradus extinction function 
                  of \citet{fitz86} is used; for the Galactic component, the
                  \citet{seat79} correction law is used.}
\tablenotetext{f}{Uncertain identification, refer to text.}
\end{deluxetable}

\clearpage
\begin{deluxetable}{lll}
\tabletypesize{\small} 
\tablewidth{0pc} 
\tablecaption{Two-shock model results} \label{tab:models}
\tablehead{
\colhead{Model}      &  \colhead{Model 1}  & \colhead{Model 2} \\
\colhead{Parameters\tablenotemark{a}} &
\colhead{$n_0 = 10^4 \pcc$}   &  \colhead{$n_0 = 3.3 \times 10^4 \pcc$}}
\startdata
  $\tau_{135}$~(years) &  0.03 $\pm$ 0.01  & \nodata  \\
  $\tau_{250}$~(years) &  2.21 $\pm$ 0.01  & \nodata  \\
  $A_{135}:A_{250}$ & \phn3.6 $\pm$ 0.7\phn & \nodata \\
  $A_{135}:A_{250}$(1)& \nodata & 5.9 $\pm$ 3.3 \\
  $A_{135}:A_{250}$(2)& \nodata & 3.3 $\pm$ 1.2 \\
  $A_{135}:A_{250}$(3)& \nodata & 2.8 $\pm$ 0.9 \\
  He                &   1 &  1 \\
  C                 &  1.85 $\pm$ 0.59 &  2.06 $\pm$ 0.66 \\
  N                 &  0.55 $\pm$ 0.08 &  0.54 $\pm$ 0.08\\
  O                 &  0.83 $\pm$ 0.16 &  0.75 $\pm$ 0.14\\
  Si                &  1.15 $\pm$ 0.37 &  0.78 $\pm$ 0.24 \\
\enddata
\tablenotetext{a}{The model parameters from top to bottom: (Model 1)
age of each shock at the first observation epoch,
ratio of shock surface areas; 
(Model 2) ratio of shock surface areas for the three 
observations (day~3869, day~4388, day~4596); 
(both Models 1 and 2)
abundances adjustment factor with respect to the ``Ring'' abundances 
(cf. \S\ref{sec-shockstruc}).
The He abundance is kept fixed for both models.}
\end{deluxetable}


\clearpage
\includegraphics[width=6in]{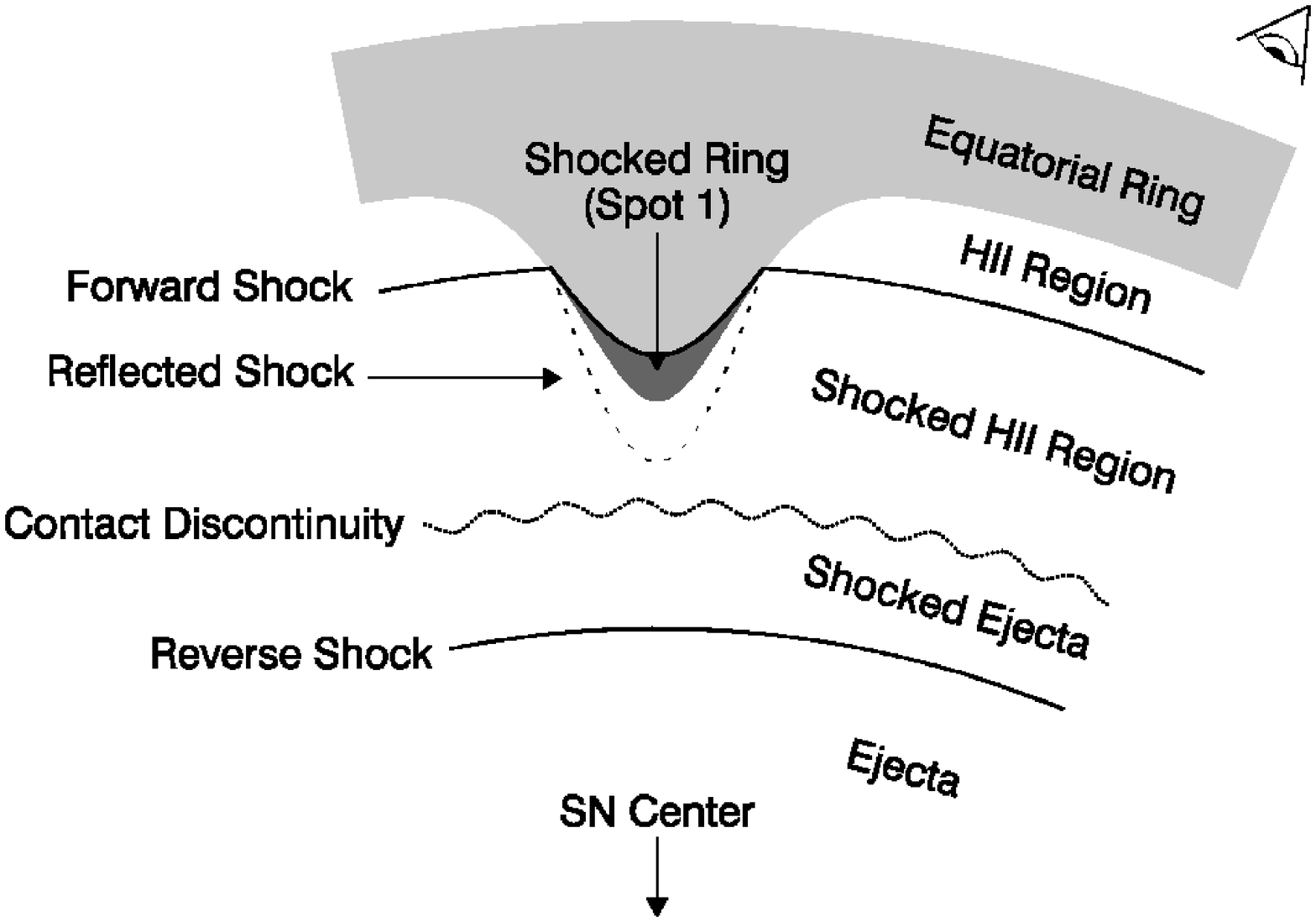}
\figcaption[f1.eps]{
Schematic representation of the double-shock structure of SNR~1987A.
\label{fig:introhydro}}

\clearpage

\clearpage

\begin{figure}
\plotone{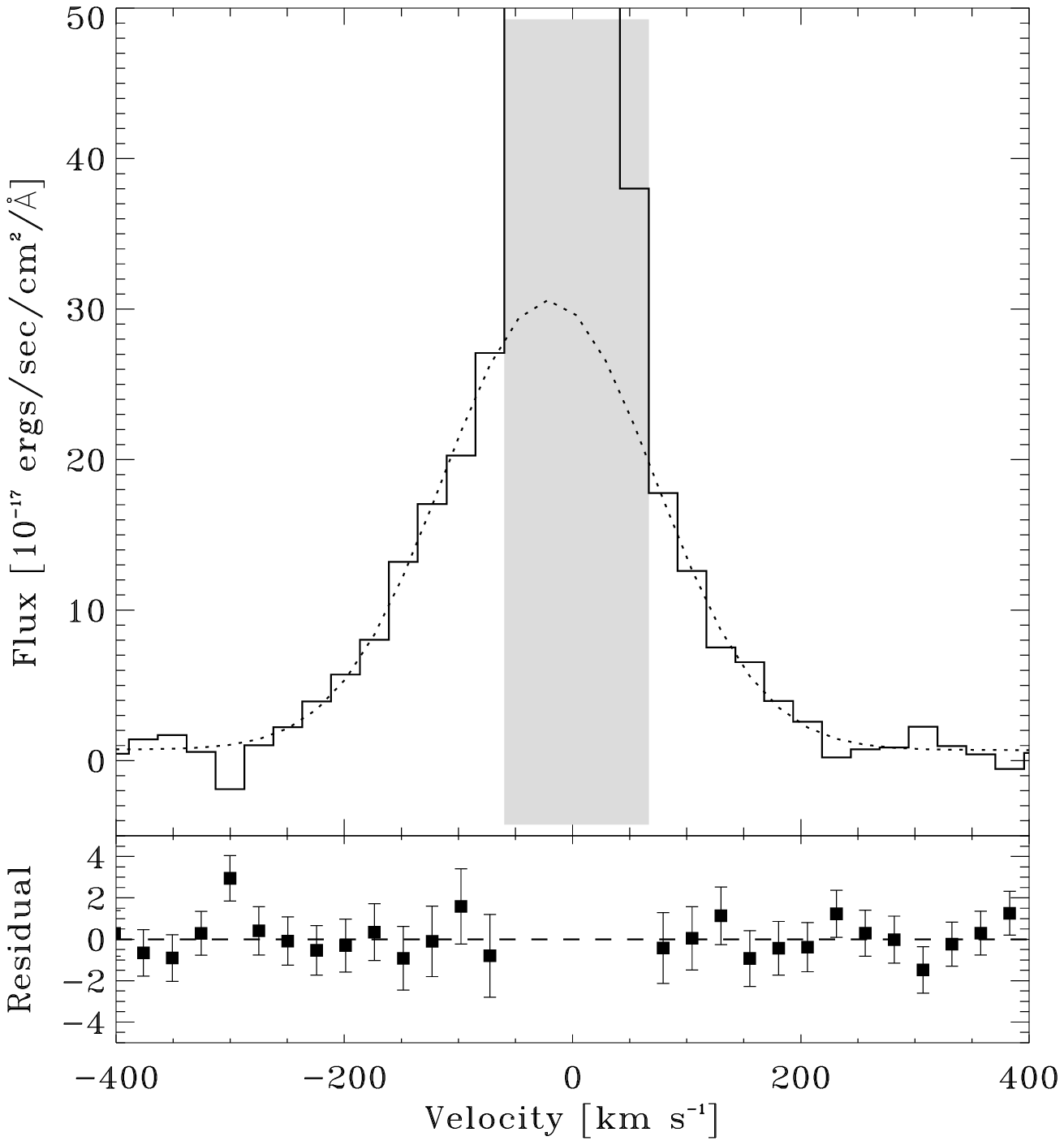}
\caption{
The profile of the [\nii] \wl6583 Spot~1 emission
obtained with the G750M (6581) grating in 1999 August.
The low-velocity region ({\it shaded})
was dominated by emission from the circumstellar ring and was excluded
from the fit. The best-fit Gaussian ({\it dotted}) to the Spot~1 emission
is shown. The velocity scale has been corrected for the heliocentric
velocity of the LMC ($285\kms$).
The residual to the best-fit Gaussian is also shown.
}
\label{fg-gaussfit}
\end{figure}

\clearpage

\clearpage

\clearpage

\clearpage

\includegraphics[width=6in]{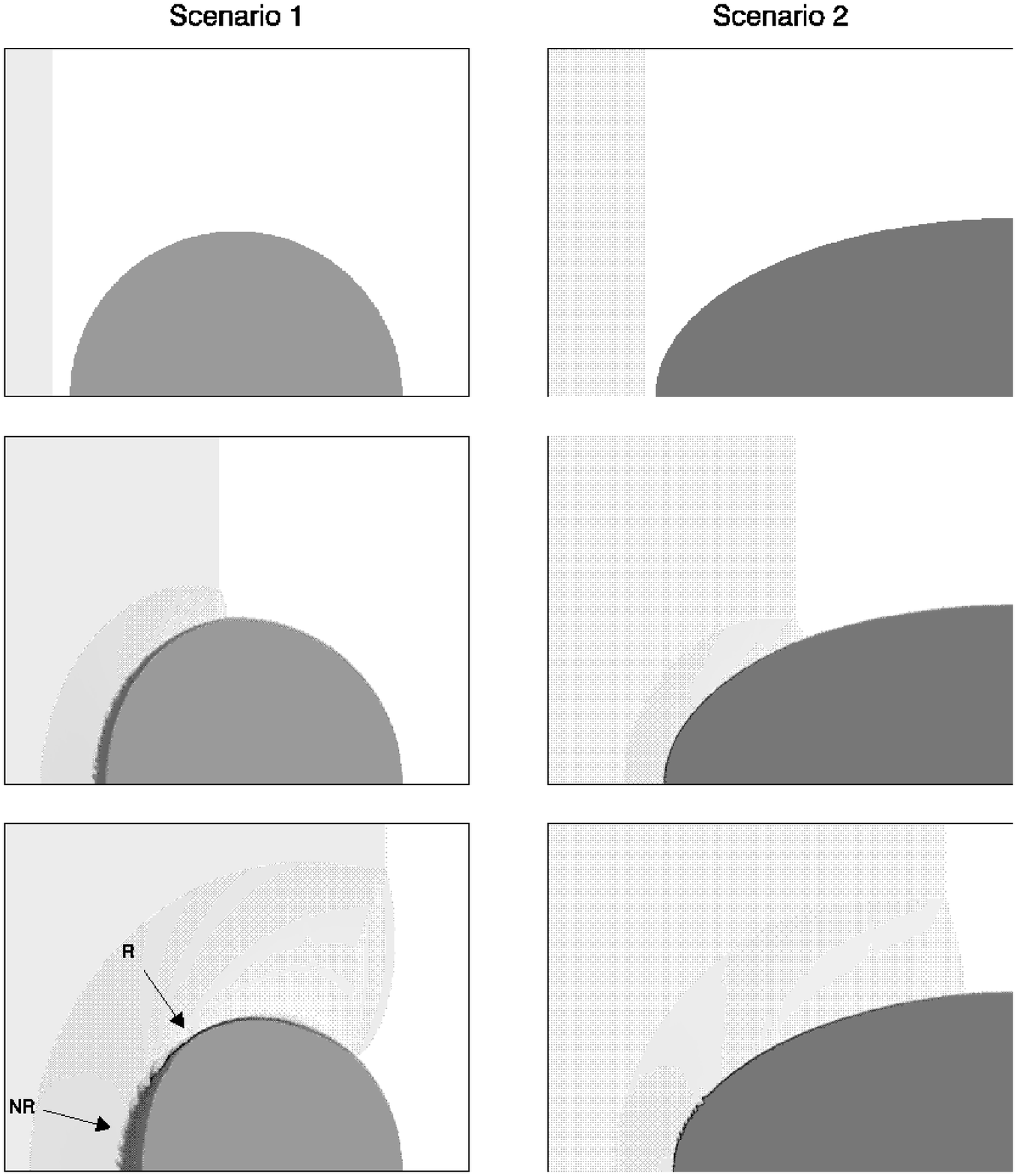}
\figcaption[f8.eps]{
2D cylindrically symmetric hydrodynamical simulations 
($800 \times 600$ zones) of a blast wave overtaking a dense 
obstacle with darker shade representing higher density.
The left panels (Scenario 1) show the development of a 
shock ($V_b = 2,000 \kms$, $\rho_{HII} = 150 \amucc$, 
$d_{spot} = 4 \times 10^{16} \cm$, $\rho_0 = 10^4 \amucc$)
where the cooling time is comparable to the 
time for the fast shock to cross the obstacle, 
i.e., $t_{cool} \approx t_{cross}$. 
Shocks that have undergone radiative collapse and developed 
dense post-shock layers (R), and shock that remained non-radiative (NR),
are indicated. 
The right panels (Scenario 2) show the development of a shock 
($V_b = 3,500 \kms$, $\rho_{HII} = 150 \amucc$,
$d_{spot} \approx 4 \times 10^{16} \cm$, $\rho_0 = 10^5 \amucc$)
where $t_{cool} \ll t_{cross}$.
Due to the high density in this obstacle, all of the
transmitted shocks undergo thermal collapse almost immediately.
\label{fig:globalhydro}}

\clearpage

\includegraphics[width=6in]{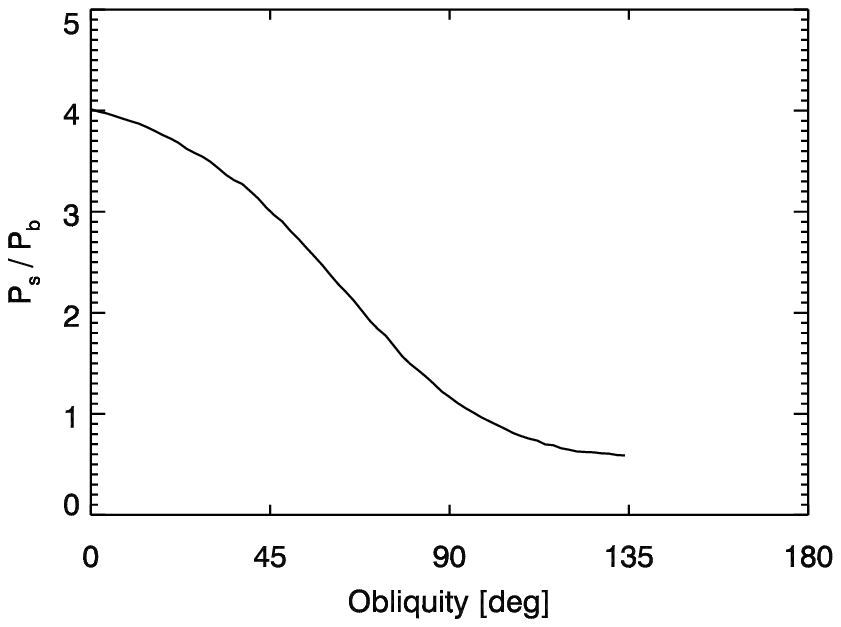} \figcaption[f9.eps]{
The pressure behind the reflected shock as a function of obliquity
immediately after impact of the blast wave for a spherical cloud having
a density enhancement $\delta = \rho_0/\rho_{HII} \approx 70$.  
We do not show the
function for obliquities higher than $135^{\circ}$ since the behavior
at the back of a cloud is complex and highly time dependent due to the
``healing'' of the blast wave.
\label{fig:beta}}

\clearpage

\includegraphics[width=6in]{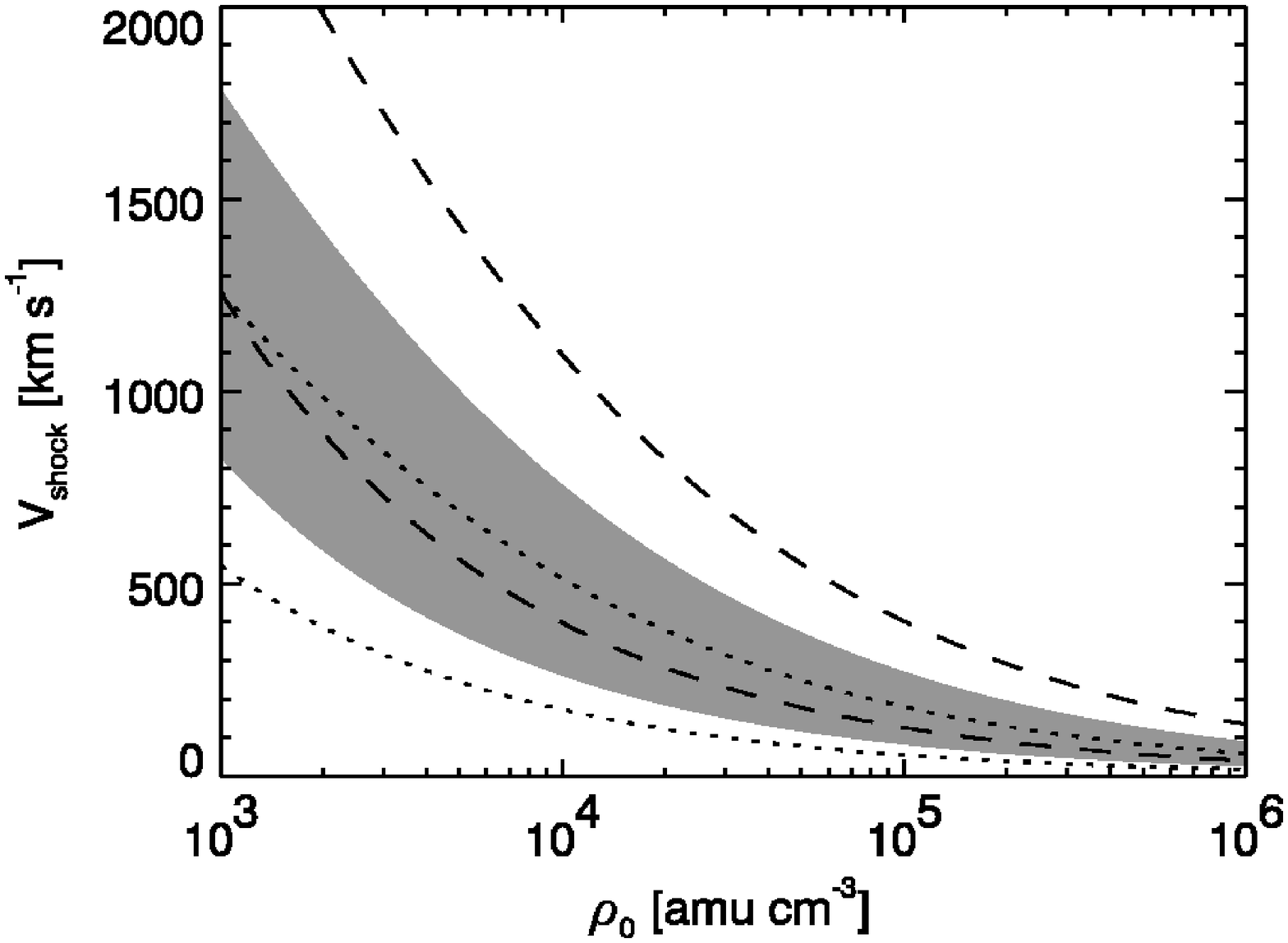}
\figcaption[f10.eps]{Range of transmitted shock velocities
($V_{shock}$) as a function of pre-shock density ($\rho_0$)
for three different estimates of the blast-wave pressure.
The shaded grey region shows the range of transmitted shock
velocities (due to obliquity) for our best estimate of the blast-wave 
pressure ($V_b = 3500 \kms$, $\rho_{HII} = 150 \amucc$).
The upper boundary is the velocity of the transmitted shock at the
tip of the protrusion and the lower boundary is the velocity at the
side of the protrusion.
The range of allowed shock velocities for our high blast-wave
pressure estimate [$V_b = 4100 \kms$, $\rho_{HII} = 250 \amucc$
({\it dashed})], and the low estimate [$V_b = 2800 \kms$, 
$\rho_{HII} = 100 \amucc$ ({\it dotted})] are also shown.
\label{fig:phasespace1}}


\includegraphics[width=5in]{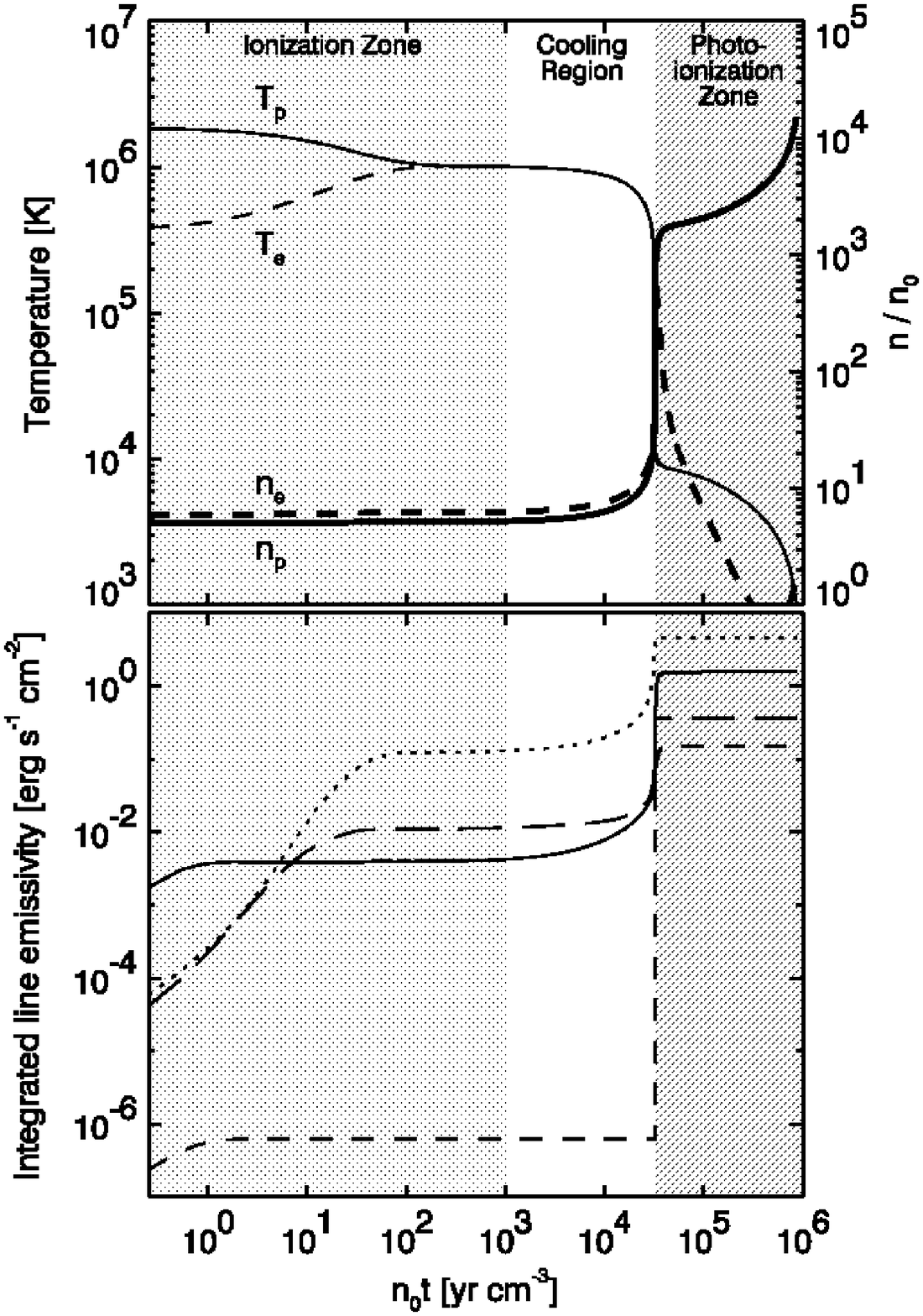} \figcaption[f11.eps]{
The top figure shows the temperature ({\it thin}) and density ({\it thick})
structure for electrons ({\it dashed}) and protons ({\it solid}) in a
radiative shock of velocity $V_s = 250 \kms$.
The bottom figure shows the integrated line emissivities for
\nv\ \wl1240 ({\it dotted}), \civ\ \wl1550 ({\it long dashes}),
[\nii] \wll 6548, 6584 ({\it short dashes}), and \Ha\ ({\it solid}).
The shading identifies the different regions of the shock as
described in the text.
\label{fig:shock}}
\clearpage

\includegraphics[width=6in]{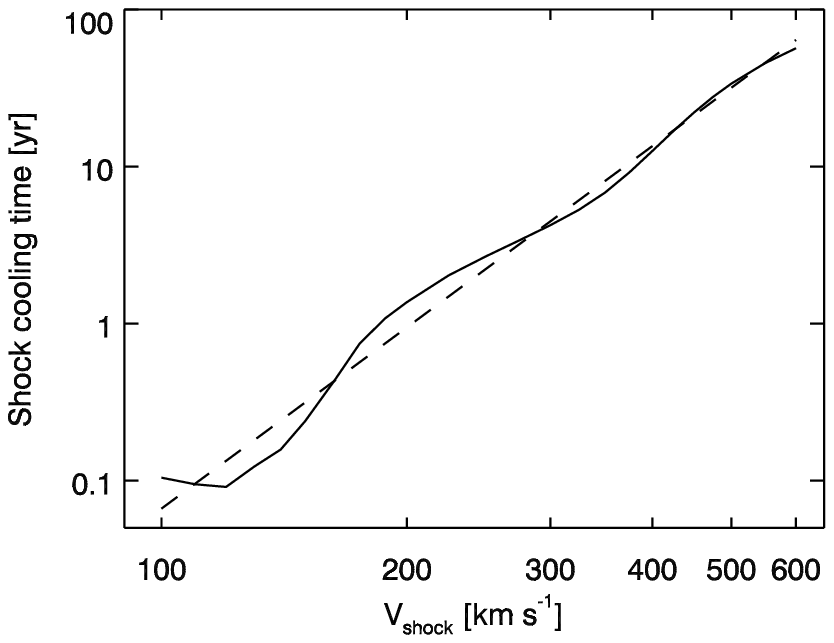}
\figcaption[f12.eps]{
Model shock cooling time $t_{cool}$ versus shock velocity $V_{shock}$
for initial obstacle density $\rho_0 = 2 \times 10^4 \amucc$ ({\it solid}).
The power-law fit $t_{cool} \propto V_s^{3.8}$ ({\it dashed}) 
is also shown.
\label{fig:cooltime}}
\clearpage

\includegraphics[width=6in]{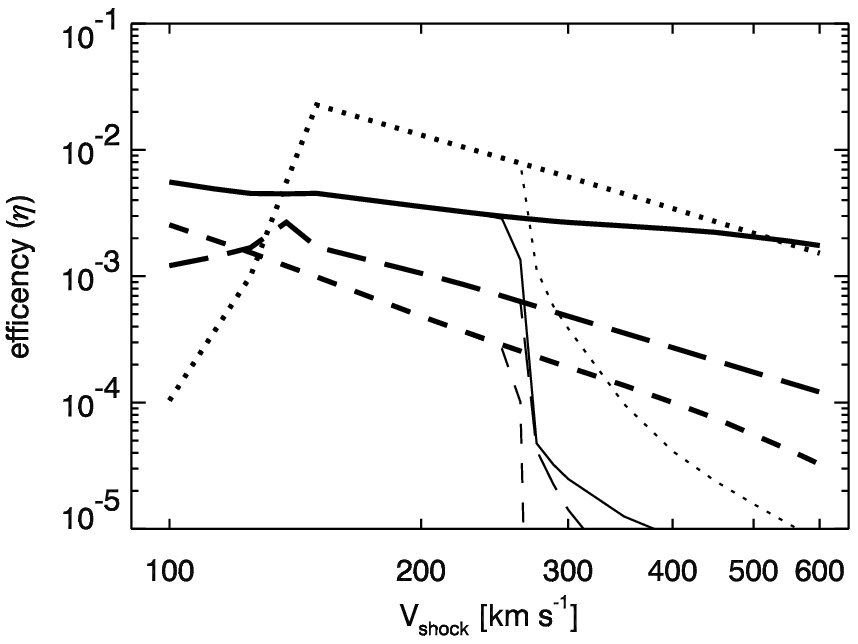} \figcaption[f13.eps]{
Line production efficiencies $\eta$ for \nv\ \wl1240 ({\it dotted}),
\civ\ \wl1550 ({\it long dashes}), [\nii] \wll6548, 6584 ({\it short dashes}),
and \Ha\ ({\it solid}) as a function of shock velocity for
fully cooled shocks ({\it thick}) and shocks which have only aged for
3 years ({\it thin}).
\label{fig:eta}}

\clearpage

\includegraphics[width=6in]{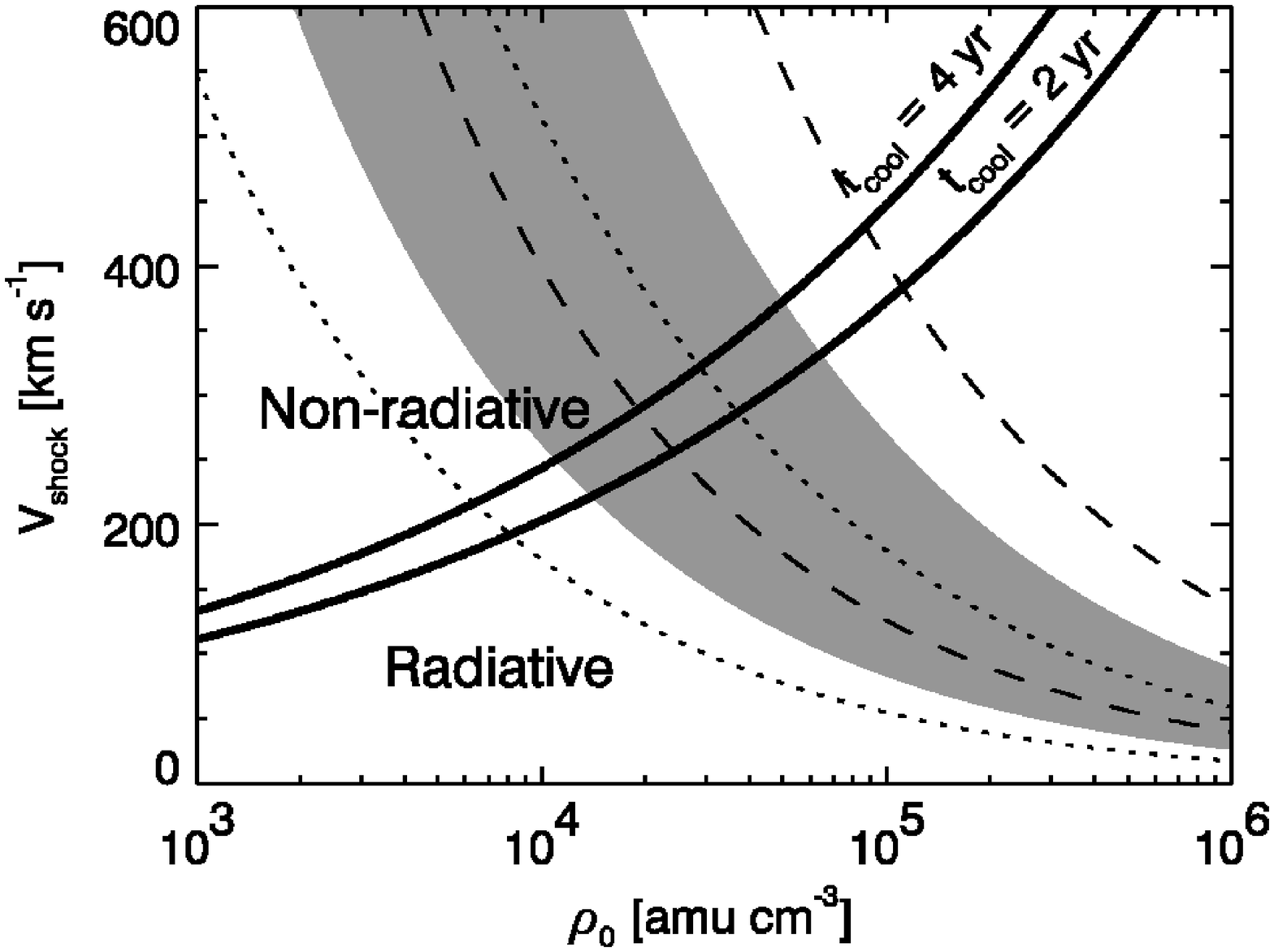}
\figcaption[f14.eps]{
Similar to Figure~\ref{fig:phasespace1}, with additional
lines ({\it thick}) dividing the phase space between radiative
shocks and non-radiative shocks for cooling time $t_{cool} =$~2 and 4~years.
\label{fig:phasespace2}}

\clearpage

\includegraphics[width=6in]{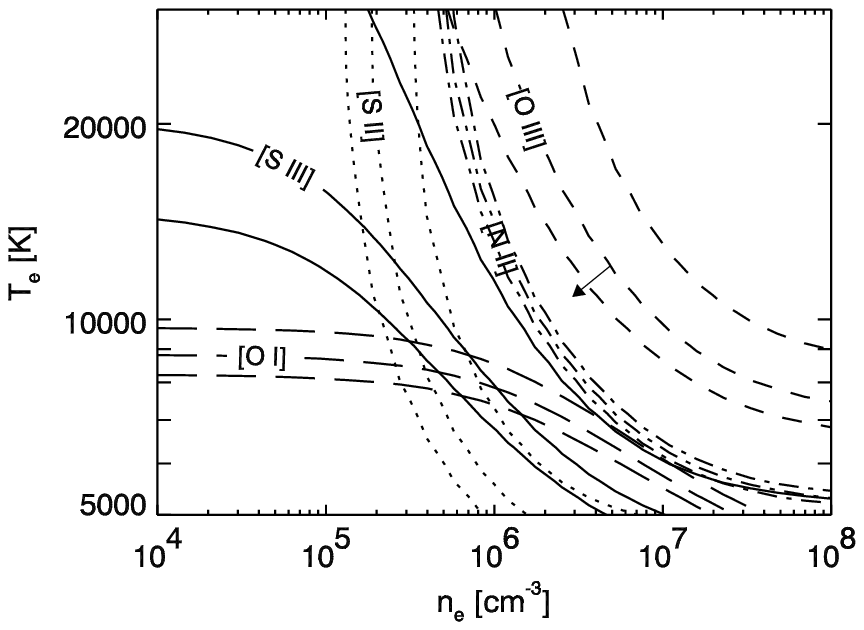} \figcaption[f15.eps]{
Contours of equal line ratios for
[\oi] (6300+6364)/5577 = $18.5 \pm 4.2$ ({\it long dash}),
[\oiii] (4959+5007)/4363 = $6.6 \pm 3.4$ ({\it short dash}),
[\nii] (6548+6584)/5755 = $2.0 \pm 0.2$ ({\it dot dash}),
[\sii] (6717+6731)/(4068+4076) = $0.1 \pm 0.04$ ({\it dotted}),
and [\siii] (9068+9530)/6312 = $20.7 \pm 10.9$ ({\it solid}).
The middle line corresponds to the determined ratio of
the dereddened line fluxes while the other two curves bracket the
1$\sigma$ errors.
Since the [\oiii] \wl4363 line may be blended with some Fe emission,
the measured ratio is a lower limit.
The arrow attached to the [\oiii] ratio contours shows
how this may affect the derived temperature and density.
\label{fig:nebanal}}

\clearpage

\includegraphics[width=5in]{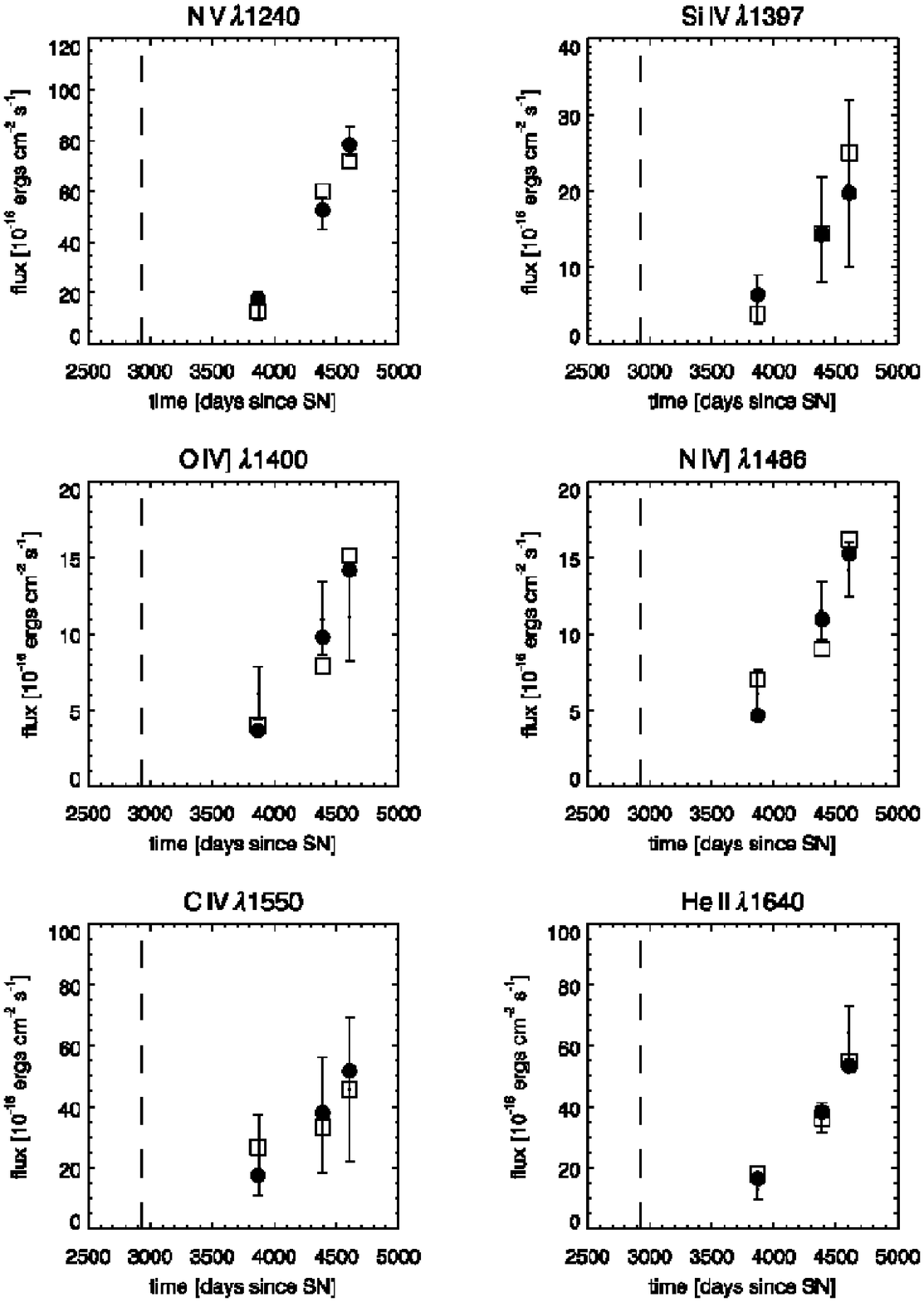} \figcaption[f16.eps]{
The observed ({\it error bars\/}) and modelled (Model 1: {\it empty squares\/};
Model 2: {\it filled circles\/}) light curves for the UV lines.
The dashed line shows the earliest date (day~2932) that Spot~1
appeared as reported by \citet{lawr00a}.
\label{fig:models}}
\clearpage

\includegraphics[width=6.in]{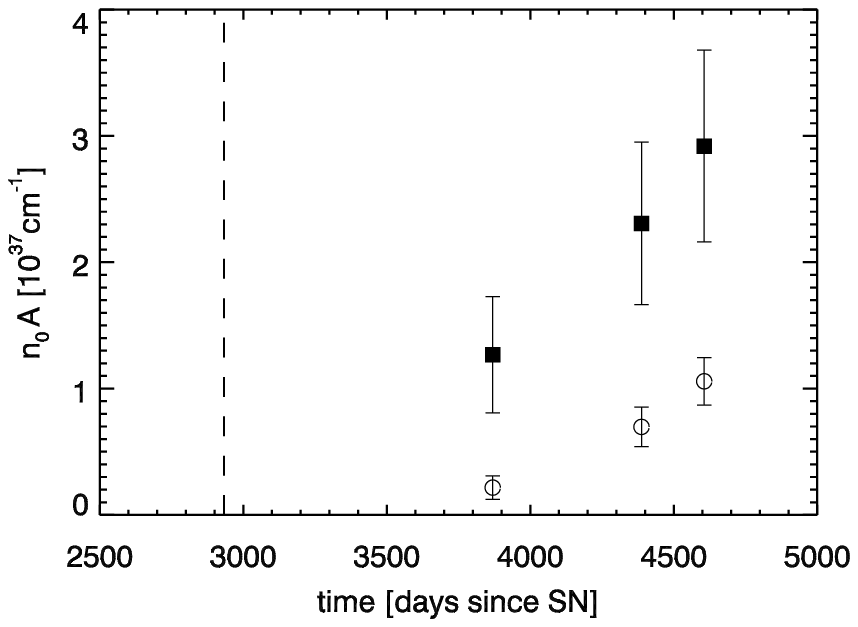} \figcaption[f17.eps]{
Shock area growth for the $V_s = 135 \kms$ ({\it filled sqaures}) 
and $V_s = 250 \kms$ ({\it empty circles}) shocks in the 
best-fit Model 2 scenario.
The dashed line shows the earliest date (day~2932) that Spot~1
appeared as reported by \citet{lawr00a}.
\label{fig:area}}
\clearpage

\includegraphics[width=6.in]{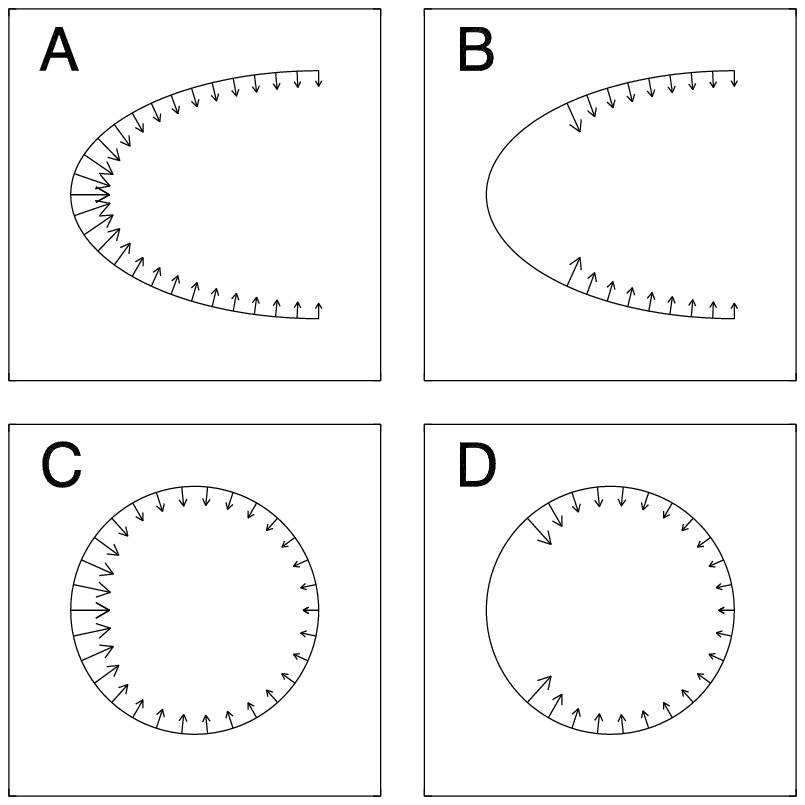}
\figcaption[f18.eps]{
Model shock-front geometries and flow 
fields for the radiative shocks in the obstacle (shock incident
from the left side).
The shock speed varies along the surface from $250 \kms$ at the
front to $100 \kms$ at the back.
Cylindrical symmetry is assumed for all models.
\label{fig:linegeom}}
\clearpage

\includegraphics[width=6in]{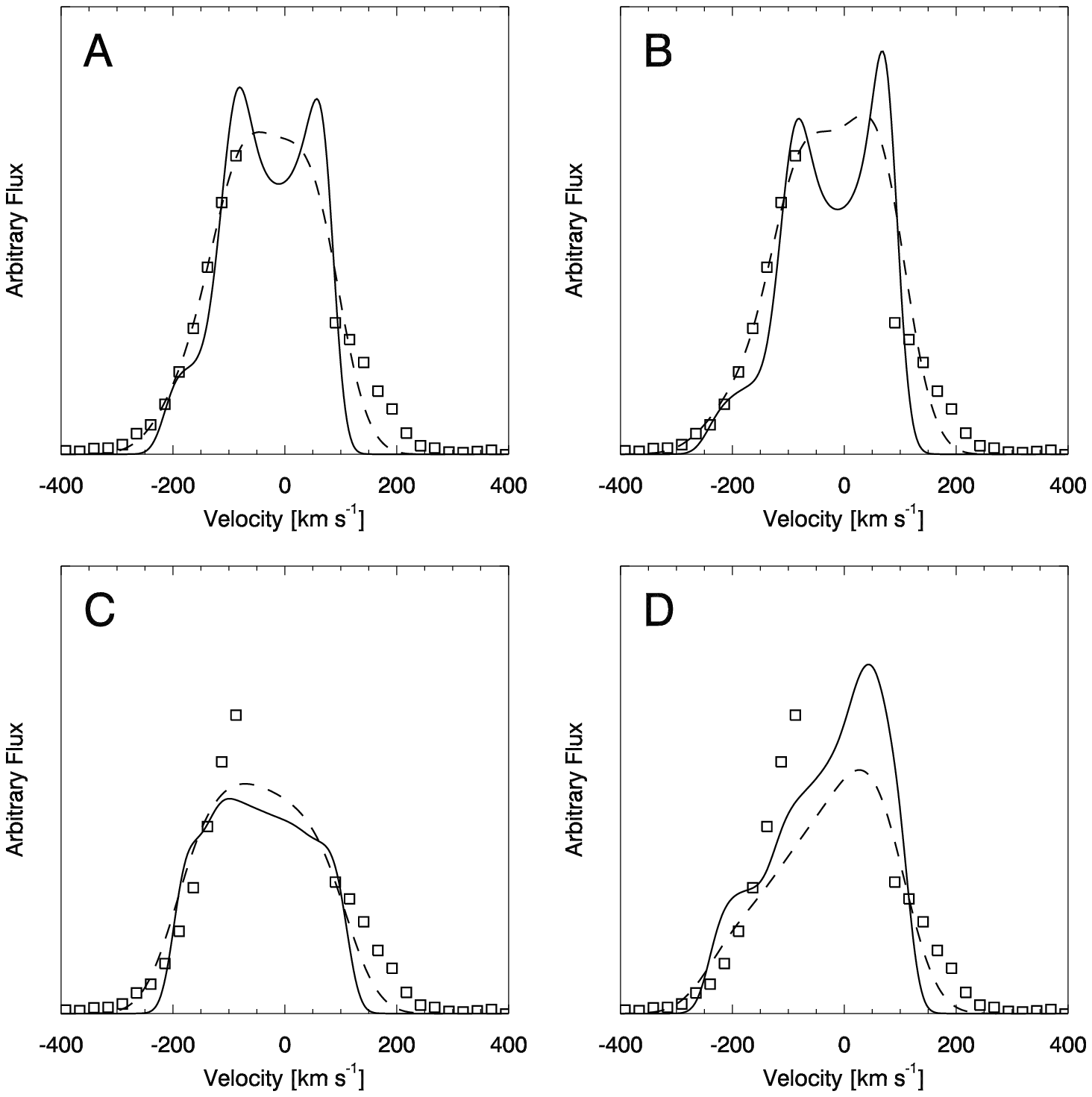}
\figcaption[f19.eps]{
Model \Ha\ line profiles ({\it solid}) for the shock-front 
geometries in Figure~\ref{fig:linegeom} are shown.
These profiles are smoothed with a Gaussian with FHWM $= 100 \kms$
({\it dotted}).
The day~4568 \Ha\ line profile from the medium resolution 0\farcs1
G750M data is plotted for comparison ({\it square}).
The uncertainties of the observed profile are smaller than the size of the
symbol.
The low-velocity region ($-50$ to $+50 \kms$) is dominated
by emission from the circumstellar ring (cf.~Figure~\ref{fg-gaussfit})
and is therefore not displayed.
\label{fig:lineprof}}
\clearpage

\includegraphics[width=6.in]{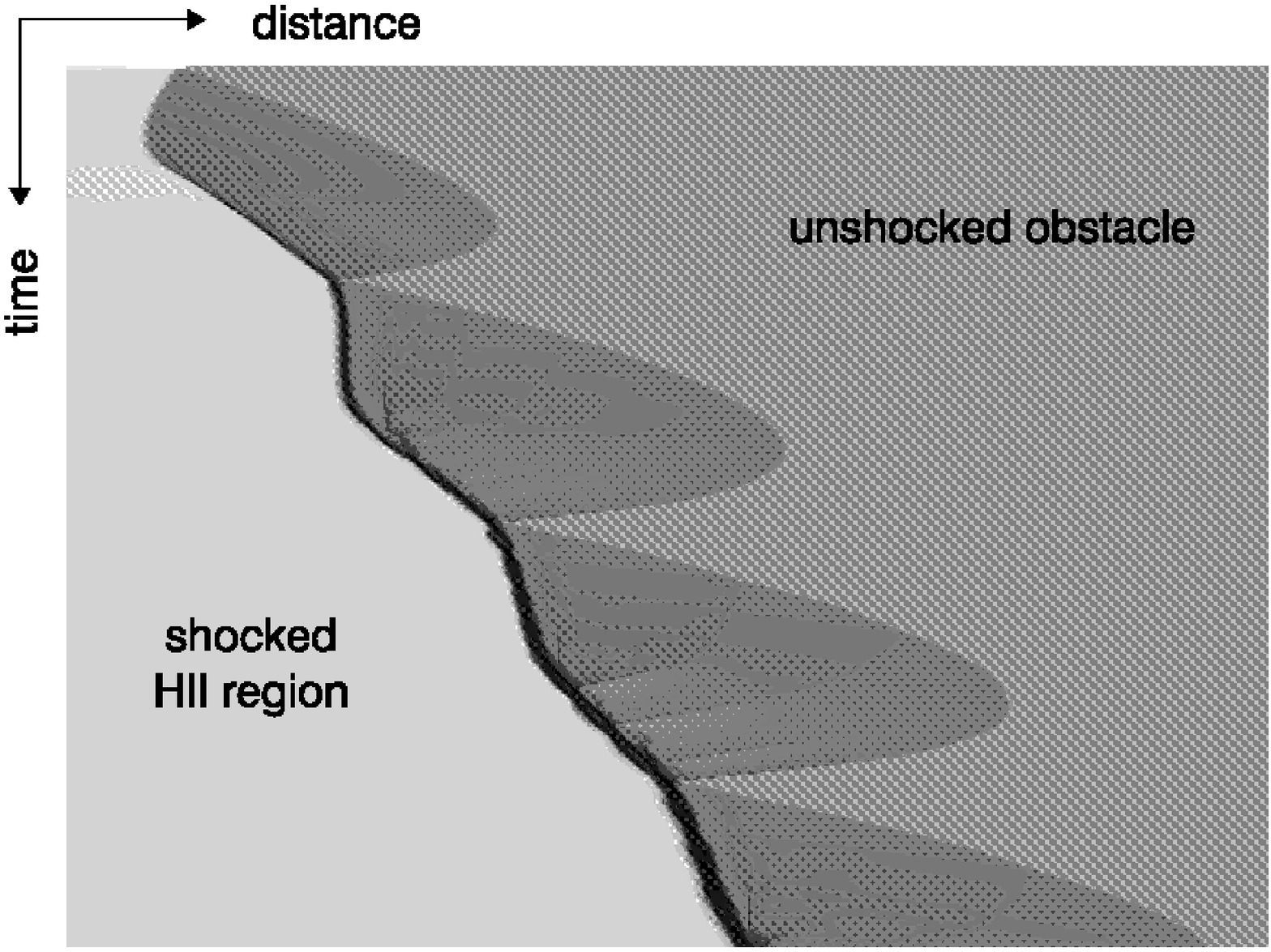} \figcaption[f20.eps]{
One-dimensional VH-1 simulation (800 spatial zones) of a radiative
shock ($V_s \approx 250 \kms$) developing in a dense ($n_0 = 3.3
\times 10^4 \pcc$) obstacle. Shown is a space-time diagram where 
darker shade represents higher density.
The overall flow is advecting to the left due to the inflow boundary
conditions chosen for the right-side boundary ($V_{inflow} = 220 \kms$).
\label{fig:therm1d}}
\clearpage

\includegraphics[width=6in]{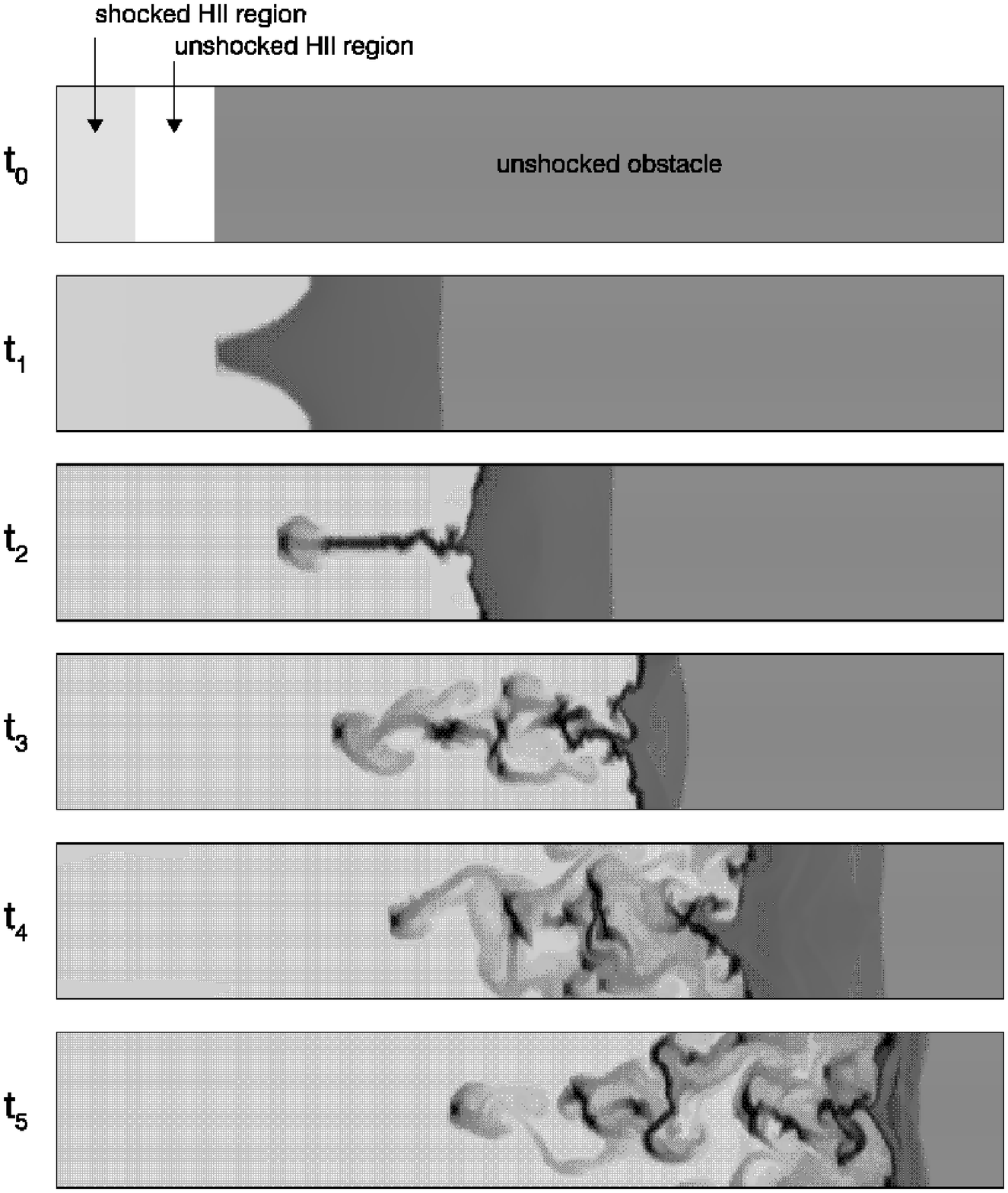} \figcaption[f21.eps]{
Two-dimensional VH-1 simulation ($80 \times 400$ zones) of the
development of a ``turbulent'' cold layer behind a radiative shock
($V_s \approx 250 \kms$) in a dense ($n_0 \approx 3.3 \times 10^4 \pcc$)
obstacle.
The elapsed time between individual frames is one year.
The obstacle has an imposed 5\% sinusoidal density
perturbation in a direction perpendicular to the shock front.
\label{fig:therm2d}
}

\end{document}